\documentclass[12pt,reqno]{amsart}
\usepackage[foot]{amsaddr}

\usepackage{hyperref,url}

\usepackage{amssymb}
\usepackage{amscd}
\usepackage{amsthm}
\usepackage{setspace}
\usepackage{enumerate}
\usepackage{graphicx}
\usepackage{mathtools}
\usepackage{tcolorbox} 
\usepackage{subfigure} 
\usepackage{siunitx}
\usepackage{tikz-cd}
\usepackage{bm,blkarray}
\numberwithin{equation}{section}

\usepackage{mathtools}
\usepackage[tableposition=top]{caption}
\usepackage{booktabs,dcolumn}

\newcommand\R{\mathbb{R}}

\newcommand\N{\mathbb{N}}
\newcommand\T{\mathbb{T}}

\newcommand{\quash}[1]{}

\renewcommand\mod[1]{\ (\mathop{\rm mod}#1)}

\renewcommand{\S}{\mathcal{S}}

\newcommand{\diag}{\mathop {\rm diag}}

\newcommand{\Hess}{\mathrm{Hess}}

\renewcommand{\mod}{\bmod}
 
\title[Average Nodal Count and the Nodal Count Condition]{Average Nodal Count and the Nodal Count Condition for Graphs}
\author{Lior Alon}
\address{Department of Mathematics, Massachusetts Institute of Technology, \newline \indent Cambridge, MA, 02139 USA.}
\email{lioralon@mit.edu}
\author{John Urschel}
\email{urschel@mit.edu}

\newtheorem{theorem}{Theorem}[section]
\newtheorem{definition}[theorem]{Definition}
\newtheorem{lemma}[theorem]{Lemma}

\newtheorem{proposition}[theorem]{Proposition}

\newtheorem{example}[theorem]{Example}

\newtheorem{corollary}[theorem]{Corollary}

\newtheorem{remark}[theorem]{Remark}
\newtheorem{ncc}[theorem]{Nodal Count Condition}

\begin{document}

\begin{abstract}
The nodal edge count of an eigenvector of the Laplacian of a graph is the number of edges on which it changes sign. This quantity extends to any real symmetric $n\times n$ matrix supported on a graph $G$ with $n$ vertices. The average nodal count, averaged over all eigenvectors of a given matrix, is known to be bounded between $\frac{n-1}{2}$ and $\frac{n-1}{2}+\beta(G)$, where $\beta(G)$ is the first Betti number of $G$ (a topological quantity), and it was believed that generically the average should be around $\frac{n-1}{2}+\beta(G)/2$. We prove that this is not the case: the average is bounded between $\frac{n-1}{2}+\beta(G)/n$ and $\frac{n-1}{2}+\beta(G)-\beta(G)/n$, and we provide graphs and matrices that attain the upper and lower bounds for any possible choice of $n$ and $\beta$. 

A natural condition on a matrix for defining the nodal count is that it has simple eigenvalues and non-vanishing eigenvectors. For any connected graph $G$, a generic real symmetric matrix supported on $G$ satisfies this nodal count condition. However, the situation for constant diagonal matrices is far more subtle. We completely characterize the graphs $G$ for which this condition is generically true, and show that if this is not the case, then any real symmetric matrix supported on $G$ with constant diagonal has a multiple eigenvalue or an eigenvector that vanishes somewhere. Finally, we discuss what can be said when this nodal count condition fails, and provide examples.

\end{abstract}

\maketitle

	\section{Introduction and Main Results}
	Let $G$ be a simple connected graph with $n$ vertices, $[n]=\{1,\ldots,n\}$, and $|E(G)|$ edges, where $E(G)$ is the edge set of $G$. The first Betti number $\beta(G)=|E(G)|-(n-1)$ of $G$ is a topological quantity measuring the maximum number of independent cycles in $G$, or, equivalently, the number of additional edges $G$ possesses over an $n$-vertex tree. We say that an $n\times n$ real symmetric matrix $A$ is \emph{supported on $G$} if $A_{ij}=0$ whenever $i\ne j$ and $(ij) \not \in E(G)$ (there are no restrictions on the diagonal of $A$), and is \emph{strictly supported on $G$} if, in addition, $A_{ij}\ne0$ for any edge $(ij) \in E(G)$. Let $\S(G)$ be the vector space of real-symmetric matrices supported on $G$, and let $\S_{0}(G)$ denote the matrices in $\S(G)$ with zeros on all diagonal entries. For example, the adjacency matrix of $G$ is in $\S_{0}(G)$, and the graph Laplacian on $G$ is in $\S(G)$ but not in $\S_{0}(G)$ (unless $G$ is regular, in which case a scalar shift of the Laplacian is in $\S_{0}(G)$). The eigenvalues of a matrix  $A\in \mathcal S(G)$ 
	are real and indexed in non-decreasing order $\lambda_1(A) \le \lambda_2(A) \le \ldots \le \lambda_n(A)$. When $\lambda_{k}(A)$ is simple, we let $\bm\varphi_{k}\in \R^{n}$ be its normalized eigenvector, and we say that $\bm\varphi_{k}$ is \emph{nowhere-vanishing} (or \emph{non-vanishing}) if $\bm\varphi_{k}(i)\ne 0$ for all $i \in [n]$. To define the \emph{nodal edge count} of $A\in\mathcal{S}(G)$, we require that:
	\begin{ncc}\label{ass}
		$A$ is strictly supported on $G$ and all its eigenvalues are simple with nowhere-vanishing eigenvectors. 
	\end{ncc} 
 \begin{remark}[NCC]
 As a convention, we use the phrase ``$A$ satisfies Nodal Count Condition \ref{ass}" in statements of theorems/propositions/lemmas, and use the abbreviation ``$A$ satisfies (NCC)" in the remainder of the text.
 \end{remark}
	If $A$ satisfies (NCC), then the nodal count sequence $\nu(A,\bm\varphi_{k})$ for $k\in[n]$ is defined by
	\[\nu(A,\bm\varphi_{k})=\left|\{i<j:A_{ij}\bm\varphi_{k}(i)\bm\varphi_{k}(j)>0\}\right|.\]
If $A_{ij}<0$ for edges $(ij)\in E(G)$, as in the case of the graph Laplacian, the nodal count is the number of edges on which $\bm\varphi_{k}$ changes sign. The study of the nodal count for matrices on graphs began with the work of Fiedler \cite{fiedler1975eigenvectors}, who, motivated by Sturm's oscillation theorem, showed that if $G$ is a tree, then $\nu(A,\bm\varphi_{k})=k-1$. Three decades later, Berkolaiko provided bounds \cite{berkolaiko2008lower}
\[k-1\le \nu(A,\bm\varphi_{k})\le k-1+\beta(G),\]
	that are tight in general. Band produced an inverse result, showing that if $\nu(A,\bm\varphi_{k})=k-1$ for all $k$, then $G$ must be a tree \cite{band2014nodal}. He did so by proving that when $\beta(G)>0$,
	\[\sum_{k=1}^{n}\left[\nu(A,\bm\varphi_{k})-(k-1)\right]>0.\] 
	Since then, the question of how the \emph{nodal surplus} $\nu(A,\bm\varphi_{k})-(k-1)$ is distributed as $k$ varies over $[n]$ was asked in the context of quantum graphs \cite{alon2018nodal}, where it was shown that the average nodal surplus is $\beta(G)/2$. For quantum graphs, numerical experiments suggest that for sufficiently large $\beta(G)$ the nodal surplus is approximately distributed like a Gaussian centered around $\beta(G)/2$ with variance of order $\beta(G)$  \cite{alon2022universality}. For matrices supported on graphs, \cite{AlonGoresky} showed that if $A\in\mathcal{S}_{0}(G)$ is fixed and the signs of its entries are changed at random, then the average nodal surplus is $\beta(G)/2$. These results led to a common belief that, in general, the average of $\nu(A,\bm\varphi_{k})-(k-1)$ for a fixed $A\in \mathcal{S}(G)$ should be $\beta(G)/2$ or close to it. 
	
\begin{figure}[t]
		\includegraphics[height=2in]{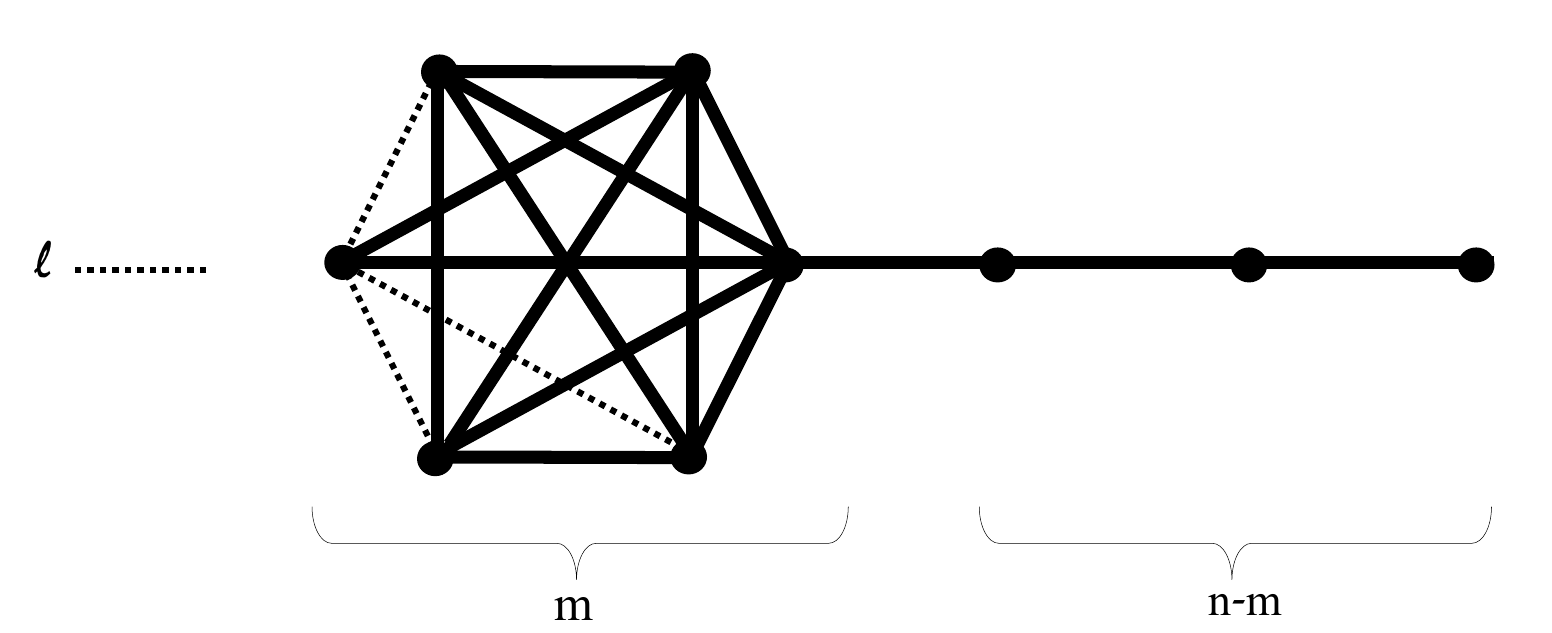} 
		\caption{An example of the extremizer graph $G=G_{n,\beta}$ obtained by attaching an $m$-clique to an $(n-m)$-vertex path, and removing $\ell$ edges from one of the clique vertices, while keeping the degree of the clique vertex connected to the path equal to $m$. The parameters $m$ and $\ell$ are taken such that ${ m-2 \choose 2} < \beta \le {m-1 \choose 2}$ and $\ell = {m -1 \choose 2} - \beta$.}
		\label{fig:minimizers}\vspace{10 mm}
	\end{figure}

	\textbf{We show that this is not the case.} Our main result is 
	\begin{theorem}\label{thm: main bounds}
		Let $G$ be a simple, connected graph $G$ on $n$ vertices, and let $A\in\mathcal{S}(G)$ satisfy Nodal Count Condition \ref{ass}. Then the average nodal surplus is bounded by
		\[\frac{\beta(G)}{n}\le\frac{1}{n}\sum_{k=1}^{n}\left[\nu(A,\bm\varphi_{k})-(k-1)\right]\le\beta(G)-\frac{\beta(G)}{n}.\]
		Equivalently, the average nodal count is bounded by 
		\[\frac{n-1}{2}+\frac{\beta(G)}{n}\le\frac{1}{n}\sum_{k=1}^{n}\nu(A,\bm\varphi_{k})\le\frac{n-1}{2}+\beta(G)-\frac{\beta(G)}{n}.\]
		Moreover, these bounds are tight.
		For any choice of $n\in\N$ and $0\le \beta\le {n-1 \choose 2}$, we construct a graph $G$ (see Figure \ref{fig:minimizers}) on $n$ vertices with $\beta(G)=\beta$ and a matrix $A\in\mathcal{S}(G)$ satisfying Nodal Count Condition \ref{ass} whose average nodal surplus attains the lower bound $\frac{\beta(G)}{n}$ (and, therefore, the average nodal surplus of $-A$ attains the upper bound).
	\end{theorem}
	\begin{remark}[lower and upper bounds are equivalent due to symmetry]\label{rem}
	If $\bm\varphi$ is the $k^{th}$ eigenvector of $A$, then it is also the $(n-k)^{th}$ eigenvector of $-A$, and $\nu(-A,\bm\varphi)=|E(G)|-\nu(A,\bm\varphi)$, where $|E(G)|=\beta(G)+n-1$ is the number of edges. As a result,  
		\[\frac{1}{n}\sum_{k=1}^{n}\nu(A,\bm\varphi_{k})=\frac{n-1}{2}+c\quad \Rightarrow\quad  \frac{1}{n}\sum_{k=1}^{n}\nu(-A,\bm\varphi_{k})=\frac{n-1}{2}+\beta(G)-c,\]
		hence the lower bound and the upper bound are equivalent.
	\end{remark}

   We prove Theorem \ref{thm: main bounds} in Subsection \ref{sub:bound} (lower and upper bounds) and Subsection \ref{sub:constructions} (tight constructions). Theorem \ref{thm: main bounds} shows that the average nodal surplus can be quite far from $\beta(G)/2$ for certain graphs $G$ (see Figure \ref{fig:minimizers}) and matrices $A \in \mathcal{S}(G)$. For zero-diagonal matrices, the same bounds obviously hold, and we can make some statements regarding their tightness. For instance, in Proposition \ref{prop:dense construction}, we prove that, for all $n \in \mathbb{N}$, there exists an $A \in S_0(K_n)$ that achieves the lower bound of Theorem \ref{thm: main bounds}, where $K_n$ is the clique on $n$ vertices. In Proposition \ref{prop: path 0diag}, we construct sparse zero-diagonal matrices that achieve the lower bound of Theorem \ref{thm: main bounds} for nodal surplus up to a multiplicative factor of two. However, in Proposition \ref{prop:biPartite}, we show that if $G$ is bipartite and $A\in\mathcal{S}_{0}(G)$ satisfies (NCC), then its average nodal surplus is exactly $\beta(G)/2$. The caveat of this result is that, while it is known that a generic matrix in $\mathcal{S}(G)$ satisfies (NCC) (see \cite[Lemma 6.1]{AlonGoreskyDisjointCycles}, for example), this is not always the case for matrices in $\mathcal{S}_{0}(G)$. Nevertheless, our second main result provides an explicit criterion for a graph $G$ so that a generic matrix in $\mathcal{S}_{0}(G)$ satisfies (NCC); if this criterion fails, then all matrices in $\mathcal{S}_{0}(G)$ fail to satisfy (NCC).
 \begin{figure}
\subfigure[Sub-Determinantal]{\includegraphics[width=5cm]{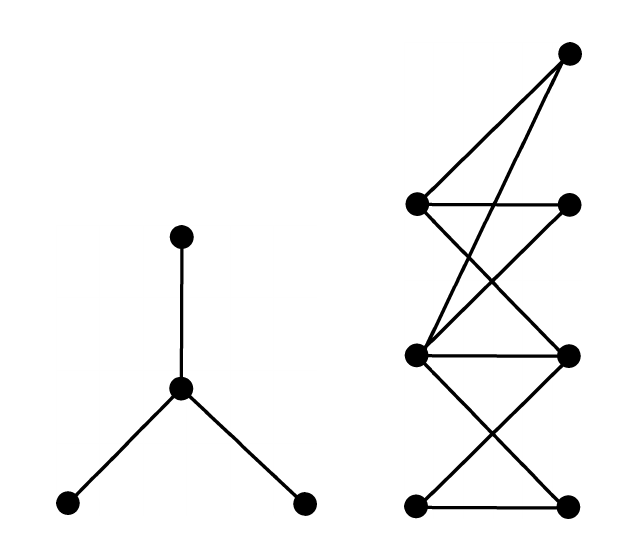}}
\hspace{2.5 cm}
\subfigure[Determinantal]{\includegraphics[width=5cm]{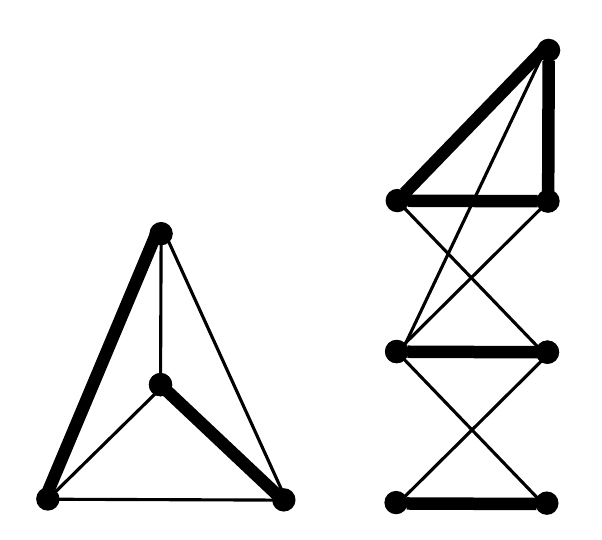}}
\caption{Examples of sub-determinantal and determinantal graphs. For each determinantal graph, we provide a vertex-disjoint union of cycles and edges covering all vertices in bold.}\label{fig: subdet}
\end{figure}

 \begin{definition}\label{def:subdeterminantal}
     We say that a graph $G$ is \emph{sub-determinantal} if $\det(A)=0$ for all $A\in \mathcal{S}_{0}(G)$, and otherwise we say that $G$ is \emph{determinantal}. Equivalently, $G$ is \emph{determinantal} if and only if it contains a vertex-disjoint union of edges and cycles that covers all vertices. For bipartite $G$, it is enough to ask if $G$ has a perfect matching.   
 \end{definition}
See Figure \ref{fig: subdet} for examples of determinantal and sub-determinantal graphs. The terminology ``sub''-determinantal suggests that if we start with $n$ vertices, and iteratively add edges one at a time, then there will be some step $j$ such that the graphs are sub-determinantal until the $j$-th step and are determinantal for all subsequent steps.  
	\begin{theorem}\label{thm: NCC characterization intro}
		Let $G$ be a simple, connected graph. The Nodal Count Condition \ref{ass} is generic in $\mathcal{S}(G)$. That is, there is an open, dense, and semi-algebraic subset of $\mathcal{S}(G)$ of matrices satisfying the condition. 
  
  For $\mathcal{S}_{0}(G)$, either $G$ is determinental, in which case the Nodal Count Condition \ref{ass} is generic in $\mathcal{S}_{0}(G)$ in a similar sense, or $G$ is sub-determinantal, in which case all matrices in $\S_{0}(G)$ fail to satisfy the condition.

  In particular, if $G$ is sub-determinantal and $A\in\mathcal{S}_{0}(G)$, then 
  \begin{enumerate}
      \item either $\lambda=0$ is a multiple eigenvalue, or
      \item the eigenvector $\bm\varphi$ of $\lambda=0$ satisfies $\bm\varphi(i)\bm\varphi(j)=0$ for every edge $(ij)$.
  \end{enumerate}
  
	\end{theorem}
The result for $\mathcal{S}_{0}(G)$ is new, and is proved in Section \ref{sec:ncc} (see Theorem \ref{thm: NCC charecterization}). The first part of the theorem, that (NCC) is generic in $\mathcal{S}(G)$, is known (see for example \cite[Lemma 6.1]{AlonGoreskyDisjointCycles}) and is only presented for comparison.  

\subsection*{Discussion: When the Nodal Count Condition fails} Finally, in Section \ref{sec:examples} we provide an analysis of the situations that can occur when (NCC) fails, either by having eigenvectors that vanish somewhere, or having multiple eigenvalues, or both (i.e., eigenvalue multiplicity and no non-vanishing orthogonal basis for the corresponding eigenspace). The theory of nodal domains on graphs is particularly focused on this setting, as when (NCC) holds, Courant's proof technique directly translates to tight results for generalized graph Laplacians. Work for nodal domains includes upper bounds for weak and strong nodal domains due to Davies, Gladwell, Leydold, and Stadler \cite{BRIANDAVIES200151,davies2000discrete}, lower bounds under (NCC) due to Berkolaiko \cite{berkolaiko2008lower}, strengthened lower bounds when (NCC) fails due to Xu and S.T. Yau \cite{xu2012nodal}, and nodal decomposition results due to the second author \cite{urschel2018nodal}. 

\begin{figure}[t]
		\includegraphics[width=6.5in]{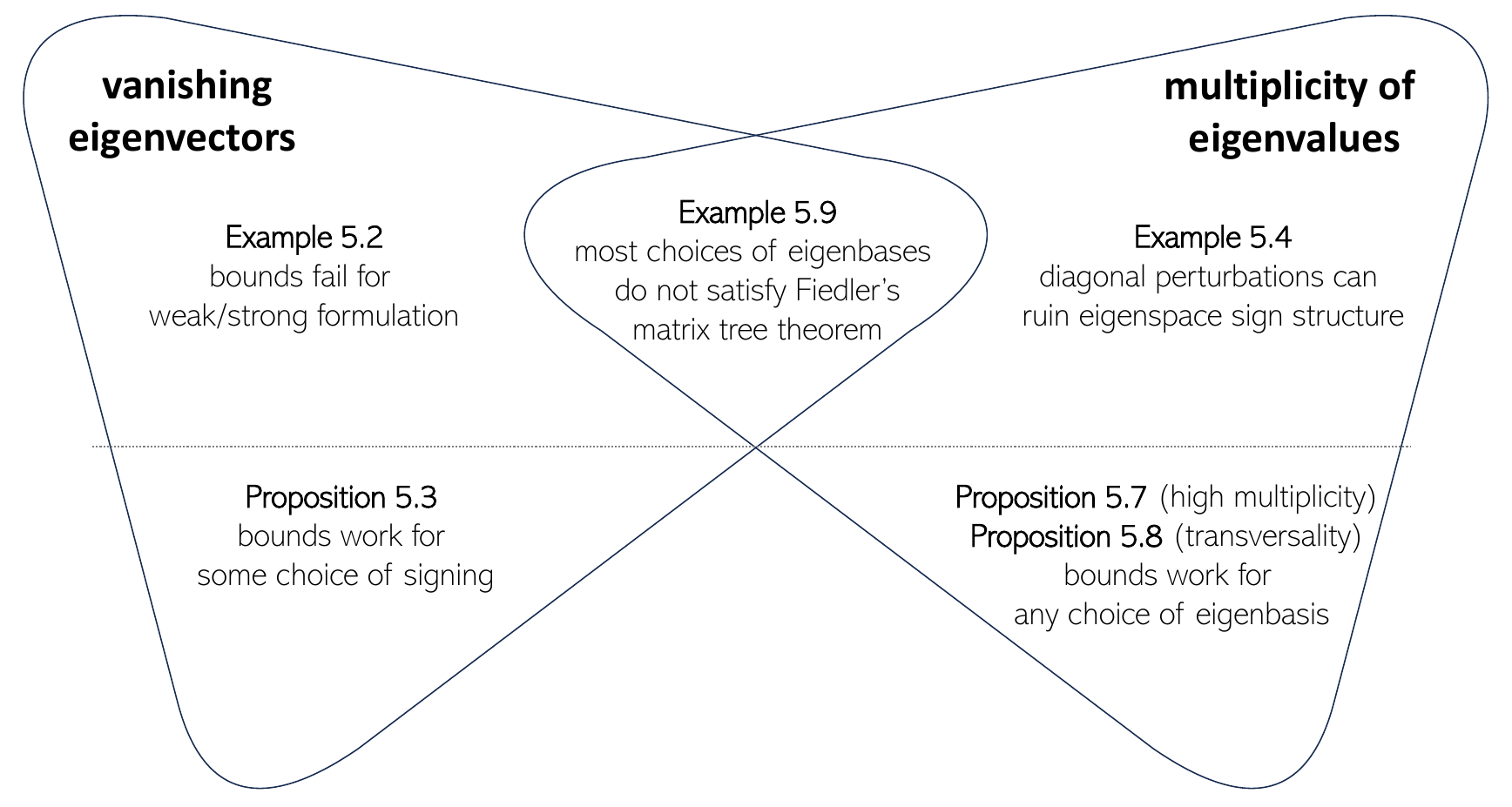} 
		\caption{A schematic diagram of what can be said when Nodal Count Condition \ref{ass} fails.}
  \vspace{10 mm}
		\label{fig:diagram}
	\end{figure}

Below we provide some results for each of these three cases for nodal count; Figure \ref{fig:diagram} summarizes our discussion into a diagram. 
 
 \begin{enumerate}
     \item \textbf{vanishing eigenvectors:} Given an eigenvector $\bm{\varphi}$ with $\bm{\varphi}(i) = 0$ for some $i$, the notion of nodal count for $\bm{\varphi}$ becomes ambiguous. We provide a construction (Example \ref{ex: vanish}) where the total nodal count, measured either with respect to edges with $A_{ij}\bm\varphi_{k}(i)\bm\varphi_{k}(j)>0$ (e.g., a strong nodal edge count) or $A_{ij}\bm\varphi_{k}(i)\bm\varphi_{k}(j)\ge 0$ (e.g., a weak nodal edge count) both fail to satisfy the bounds of Theorem \ref{thm: main bounds}. However, in Proposition \ref{prop: signing} we show that there always exists a signing of the vanishing entries of eigenvectors, so that the resulting augmented vectors have a total nodal count satisfying the bounds of Theorem \ref{thm: main bounds}. \\
     \item \textbf{eigenvalue multiplicity:} When a matrix has a repeated eigenvalue, there is no longer a notion of a fixed eigenbasis. A naive approach would be to simply choose a nowhere-vanishing eigenbasis and hope that a small enough perturbation would resolve eigenvalue multiplicity without changing the sign structure, leading back to the Theorem \ref{thm: main bounds} bounds. We provide an example where this type of approach fails (Example \ref{ex: cycle}). In particular, we produce an example where the sign structure of the eigenbasis of the perturbed matrix does not match the sign structure of any eigenbasis of the original matrix. On the other hand, suppose that $A\in\mathcal{S}(G)$ has $r$ distinct eigenvalues of multiplicities $(m_{i})_{i=1}^{r}$. Proposition \ref{prop: multiplicity} shows that if $\sum_{i=1}^{r}{m-1 \choose 2}\ge \beta(G)$, then any nowhere-vanishing eigenbasis of $A$ satisfies the bounds of Theorem \ref{thm: main bounds}, and Proposition \ref{prop: trans-mult} shows that for the case $\sum_{i=1}^{r}{m-1 \choose 2}< \beta(G)$, if 
$A$ satisfies a certain transversality condition, then any nowhere-vanishing eigenbasis of $A$ satisfies the bounds of Theorem \ref{thm: main bounds}.\\
    \item \textbf{eigenvalue multiplicity \& vanishing eigenvectors:} When all the eigenvectors of a repeated eigenvalue vanish at a vertex, the behavior of eigenvectors becomes rather complex. The adjacency matrix $A$ of $G$, a star graph on $n$ vertices (Example \ref{ex: star}), is a standard example illustrating the difficulties in this case. For this graph, the relative size of the subset of eigenbases of $A$ that can be perturbed to an eigenbasis of a nearby matrix $A' \in \S(G)$ satisfying (NCC) is exponentially small in $n$.
 \end{enumerate}

	\section{Preliminary Results: perturbations, transversality, matrix inversion, and the nodal-magnetic relation}\label{sec:tech}
	\subsection{Additive perturbation and signs of eigenvectors}  
 Here we collect some useful tools from perturbation theory. We mainly concern ourselves with the simple setting $A_{\epsilon}=A+\epsilon B$, however, we sometimes require the more general setting of a one-parameter real analytic family $A_{\epsilon}$ of real symmetric matrices. We start with a celebrated result proven by Kato \cite[Thm 1.8]{kato2013perturbation} and by Rellich \cite[Thm. 1]{rellich1969perturbation}, see also \cite{wimmer1986rellich} for a short proof.
 \begin{lemma}\cite{kato2013perturbation,rellich1969perturbation}\label{lm: katto-rellich}
 A one-parameter family of real symmetric matrices $A_{\epsilon}$ that depends analytically on $\epsilon$ admits an orthonormal basis of eigenvectors that depends analytically on $\epsilon$. 
 \end{lemma}
 The following lemma is immediate. 
\begin{lemma}\label{lm: perturbation general} Let $A_{\epsilon}$ be a one-parameter real analytic family of $n\times n$ real symmetric matrices, let $\bm\varphi_{1,\epsilon},\bm\varphi_{2,\epsilon},\ldots,\bm\varphi_{n,\epsilon}  $ be a real analytic orthonormal eigenbasis, with (not necessarily ordered) real eigenvalues $\lambda_{k,\epsilon}=\langle A_{\epsilon}\bm\varphi_{k,\epsilon},\bm\varphi_{k,\epsilon} \rangle$. The power series expansions $A_{\epsilon}=A^{(0)}+\epsilon A^{(1)}+O(\epsilon^{2}),\ \lambda_{k,\epsilon}=\lambda_{k}^{(0)}+\epsilon \lambda_{k}^{(1)}+O(\epsilon^{2})$, and $\bm\varphi_{k,\epsilon}=\sum_{j=0}^{\infty}\epsilon^{j}\bm\varphi_{k}^{(j)}$, satisfy   
 \begin{enumerate}
     \item we have $\lambda_{k}^{(1)}=\langle A^{(1)}\bm\varphi_{k}^{(0)},\bm\varphi_{k}^{(0)}\rangle$, and $\langle A^{(1)}\bm\varphi_{k}^{(0)},\bm\varphi_{k'}^{(0)}\rangle=0$ whenever $\lambda_{k}^{(0)}=\lambda_{k'}^{(0)}$ with $k'\ne k$. In particular, the restriction of $A^{(1)}$ to the eigenspace $\ker(A^{(0)}-\lambda_{k}^{(0)})$ is diagonal in the basis of $\bm\varphi_{k'}^{(0)}$ with $\lambda_{k'}^{(0)}=\lambda_{k}^{(0)}$.
     \item There is a discrete (locally finite) set $\mathcal{B}\subset \R$ such that for all $\epsilon\in \R\setminus \mathcal{B}$ the support of $\bm\varphi_{k,\epsilon}$ is independent of $\epsilon$ and given by 
     $$\mathrm{supp}(\bm\varphi_{k,\epsilon})=\bigcup_{j=0}^{\infty}\mathrm{supp}(\bm\varphi_{k}^{(j)}).$$
     \item Let $b>0$ such that $(0,b)\cap\mathcal{B}=\emptyset$, let $i\in \bigcup_{j=0}^{\infty}\mathrm{supp}(\bm\varphi_{k}^{(j)})$ and let $j_{i}$ be the minimal $j$ for which $\varphi_{k}^{(j)}(i)\ne 0$, then for all $\epsilon\in (0,b)$,
     		\begin{equation}\label{eq: sign veps}
				\mathrm{sign}(\bm\varphi_{k,\epsilon}(i))=\mathrm{sign}(\bm\varphi_{k}^{(j_{i})}(i)).  
			\end{equation} 
 \end{enumerate} 
 \end{lemma}
 \begin{proof}
     Part (1) is given by the first-order term in the expansion of $\langle (A_{\epsilon}-\lambda_{k,\epsilon} )\bm\varphi_{k,\epsilon},\bm\varphi_{k',\epsilon}\rangle\equiv 0 $. Part (2) is due to the fact that for each $i\in[n]$, either $\bm\varphi_{k}^{(j)}(i)=0$ for all $j$, or the real analytic function $\epsilon\mapsto \bm\varphi_{k,\epsilon}(i)=\sum_{j=0}^{\infty}\epsilon^{j}\bm\varphi_{k}^{(j)}(i)$ has a discrete set of zeros. For part (3) notice that $\bm\varphi_{k,\epsilon}(i)\ne 0$ for all $\epsilon\in (0,b)$, since $(0,b)\cap\mathcal{B}=\emptyset$, so $\mathrm{sign}(\bm\varphi_{k,\epsilon}(i))$ is constant in that interval. For small enough $\epsilon>0$ we have 
     $\bm\varphi_{k,\epsilon}=\epsilon^{j_{i}}(\bm\varphi_{k}^{(j_{i})}(i)+O(\epsilon))$ which determine the sign. 
 \end{proof}
Now consider the simple case of $A_{\epsilon}=A+\epsilon B$.
\begin{lemma}\label{lm: perturbation simple} Let $A_{\epsilon}=A+\epsilon B$ with $A$ and $B$ real symmetric $n\times n$ matrices, then 
\begin{enumerate}
    \item If $A$ has simple eigenvalues, then $A_{\epsilon}$ has simple eigenvalues for all $\epsilon\in(-T,T)$ with 
    $$T=\frac{\min_{k}|\lambda_{k+1}(A)-\lambda_{k}(A)|}{\max_{k}|\lambda_{k}(B)|}.$$
    \item Let $\bm\varphi_{\epsilon}=\sum_{j=0}^{\infty}\epsilon^{j}\bm\varphi^{(j)}$ be an eigenvector of $A_{\epsilon}$ with a simple eigenvalue for all $\epsilon$ in some interval $(a,b)$. Then in this interval, we have a recursive formula
			\begin{equation}\label{eq: pert vj full}
				\bm\varphi^{(j)}=-(A-\lambda^{(0)} I)^{+}B\bm\varphi^{(j-1)}+\sum_{m=1}^{j}\lambda^{(m)}(A-\lambda^{(0)}I)^{+}\bm\varphi^{(j-m)},  
			\end{equation}
   where $(A-\lambda^{(0)}I)^{+}$ denotes the pseudo-inverse\footnote{The pseudo-inverse $M^{+}$ of a real symmetric matrix $M$ is a real symmetric matrix with the same kernel of $M$, such that $M^{+}M=MM^{+}$ acts as the identity on the orthogonal complement of the kernel.} of $A-\lambda^{(0)}I$. This formula becomes simpler, if we assume that $\lambda^{(0)}=0$ and $\langle B \bm\varphi^{(0)}, \bm\varphi^{(0)}\rangle=0$, in which case $\bm\varphi^{(1)}=-A^{+}B\bm\varphi^{(0)}$ and, for all $j>1$,		\begin{equation}\label{eq: pert vj}
				\bm\varphi^{(j)}=-A^{+}B\bm\varphi^{(j-1)}+\sum_{m=2}^{j}\lambda^{(m)}A^{+}\bm\varphi^{(j-m)}.  
			\end{equation}
   \item Let $\bm\varphi_{1,\epsilon},\ldots,\bm\varphi_{n,\epsilon}$ be the orthonormal eigenvectors of $A_{\epsilon}$ as in the previous lemma, and suppose that $\lambda_{k,0}=\lambda$ is a multiple eigenvalue of $A$ and that $\lambda_{k,\epsilon}$ is simple for all $\epsilon\in (0,b)$ for some $b>0$. Then,  $\lambda_{k}^{(1)}\ne \lambda_{k'}^{(1)}$ for any  $ k'\ne k$ with $\lambda_{k'}^{(0)}=\lambda$, and the following holds for all $\epsilon\in (0,b)$, 
   \begin{equation}\label{eq: pert vj mult}
       \bm\varphi_{k}^{(1)}  =-(A-\lambda)^{+}B\bm\varphi_{k}^{(0)}+\sum_{ \lambda_{k'}^{(0)}=\lambda,\ k'\ne k}\frac{\langle (A-\lambda)^{+}B\bm\varphi_{k'}^{(0)},B\bm\varphi_{k}^{(0)}\rangle}{\lambda_{k}^{(1)}-\lambda_{k'}^{(1)}}\bm\varphi_{k'}^{(0)}. 
   \end{equation}
\end{enumerate}     
 \end{lemma}
\begin{proof}
    For part (1) see \cite[Cor. 4.3.15]{horn2012matrix} for the value of $T$ that ensures simple eigenvalues when $\epsilon \in (-T,T)$. 
    For part (2), the eigenvalue-eigenvector equation $(A+\epsilon B-\lambda_{\epsilon}I)\bm\varphi_{\epsilon}=0$ decomposes according to powers of $\epsilon$, with the below equation representing $\epsilon^j$, as follows:
  \[(A-\lambda^{(0)} I)\bm\varphi^{(j)}=  -B\bm\varphi^{(j-1)}+\sum_{i=1}^{j}\lambda^{(i)}\bm\varphi^{(j-i)}.\]
  the normalization condition $\langle \varphi_{\epsilon}, \varphi_{\epsilon}\rangle\equiv 1$ decomposes according to powers of $\epsilon$ gives that $\langle \varphi^{(0)}, \varphi^{(j)}\rangle=0$ for all $j>0$, and using the assumption that $\lambda^{(0)}$ is simple we get that $\varphi^{(j)}$ is in the range of $(A-\lambda^{(0)}I)^{+}$. Equation \eqref{eq: pert vj full} is obtained by multiplying both sides by $(A-\lambda^{(0)}I)^{+}$.
  
  For Equation \eqref{eq: pert vj}, note that the inner product of the $j=1$ equation with $\bm\varphi^{(0)}$ gives $\lambda^{(1)} =\langle \bm\varphi^{(0)},B\bm\varphi^{(0)}\rangle$, so the $j=1$ equation gives $A\bm\varphi^{(1)}=-B\bm\varphi^{(0)}$, and so Equation \eqref{eq: pert vj} follows.

  Finally, for part (3), let us write 
  $\bm\varphi_{k}^{(1)}=w+\sum_{ \lambda_{k'}^{(0)}=\lambda,\ k'\ne k}c_{k,k'}\bm\varphi_{k'}^{(0)}$, for some scalars $c_{k,k'}$ and a vector $w\perp \ker(A-\lambda I)$, using that $\langle \bm\varphi_{k}^{(1)},\bm\varphi_{k}^{(0)}\rangle=0$. The $j=1$ term of the eigenvalue-eigenvector equation gives 
  \begin{align*}
     (A-\lambda I)w & =(A-\lambda I)\bm\varphi_{k}^{(1)}=  -B\bm\varphi_{k}^{0}+\lambda^{(1)}\bm\varphi^{0},\ \text{so}\\ 
     w & =-(A-\lambda I)^{+}B\bm\varphi_{k}^{0}.
  \end{align*}

   To find $c_{k,k'}$ we take the inner product of $\bm\varphi_{k'}^{(0)}$ and the second-term of the eigenvalue-eigenvector equation 
   \[(A-\lambda_{k}^{(0)} I)\bm\varphi_{k}^{(2)}=  -B\bm\varphi_{k}^{1}+\lambda_{k}^{(1)}\bm\varphi_{k}^{1}+\lambda_{k}^{(2)}\bm\varphi_{k}^{0},\]
where the right-hand-side is equal to $-Bw+c_{k,k'}(\lambda_{k}^{(1)}-B)\bm\varphi_{k'}^{(0)}$ plus a vector orthogonal to $\bm\varphi_{k'}^{(0)}$, so  $$c_{k,k'}=\frac{\langle -Bw, \bm\varphi_{k'}^{(0)}\rangle}{\langle (\lambda_{k}^{(1)}-B)\bm\varphi_{k'}^{(0)}, \bm\varphi_{k'}^{(0)}\rangle}= \frac{\langle -Bw, \bm\varphi_{k'}^{(0)}\rangle}{\lambda_{k}^{(1)}-\lambda_{k'}^{(1)}}.$$
	\end{proof}

	\subsection{Transversality}
\begin{figure}[t]
		\includegraphics[width=3.5in]{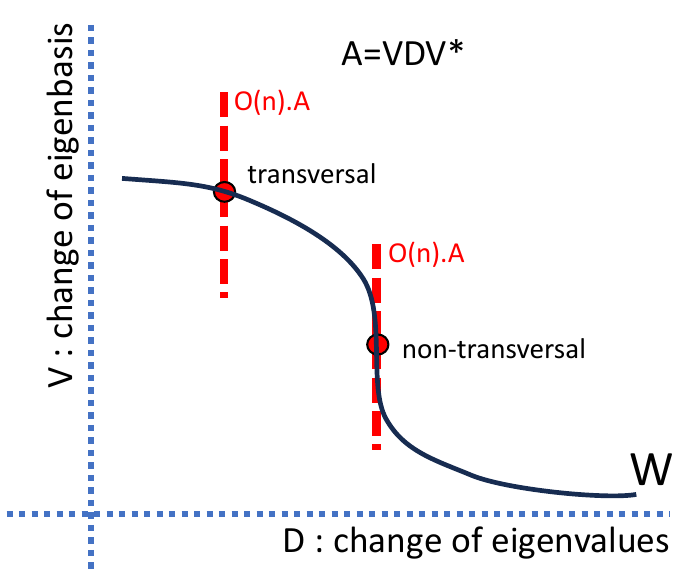} 
		\caption{A heuristic diagram to illustrate transversality of $O(n)$ orbits with respect to a vector space $W\subset\mathcal{S}(n)$. The coordinate system, $D$ against $V$ (in the spectral decomposition $A=VDV^{T}$), means that vertical lines are $O(n)$ orbits and horizontal lines contain all matrices with the same eigenbasis. }\vspace{10 mm}
		\label{fig:trans}
	\end{figure}

	When a matrix $A\in\mathcal{S}(G)$ has eigenvalue multiplicity, there is no longer a unique choice of eigenbasis, and different choices can lead to different nodal counts. One naive approach to bound nodal counts in this situation would be to slightly change the eigenvalues to resolve multiplicity, and then slightly change the eigenvectors to ensure the resulting matrix $A'$ is in $\mathcal{S}(G)$, all while maintaining the sign structure of the matrix and eigenbasis. The orbit of a matrix $A\in\mathcal{S}(G)$ under the action of the orthogonal group $O(n)$, which we denote by $O(n).A:=\{oAo^{T}:o\in O(n)\}$, is the set of matrices (not necessarily in $\mathcal{S}(G)$) with the same spectrum. A sufficient condition for this naive approach to work is that the intersection of $O(n).A$ and $\mathcal{S}(G)$ at $A$ is transversal (see Figure \ref{fig:trans}). Namely, this transversality is a sufficient condition so that for any choice of nowhere-vanishing eigenbasis of $A$ there is a nearby matrix $A'\in\mathcal{S}(G)$ satisfying (NCC) whose eigenbasis has the same sign structure and, therefore, the same nodal count. We state the following result for a general vector space $W$, which will be applied to either $W=\mathcal{S}(G)$ or $W=\mathcal{S}_{0}(G)$. 
	\begin{lemma}[Transversality]\label{lem: transversality} Let $\mathcal{S}(n)$ and $\mathcal{A}(n)$ be the set of real symmetric and real anti-symmetric matrices, respectively, $A\in \mathcal{S}(n)$, and $O(n).A=\{oAo^{T}: o\in O(n) \}$.
 \begin{enumerate}
     \item The tangent space at $A$ is $T_{A}(O(n).A)=\{AX-XA: X\in\mathcal{A}(n)\}$. If $A$ has $r$ distinct eigenvalues of multiplicities $m_1,\ldots,m_r$, $\sum_{i=1}^r m_i = n$, then
     \[\dim(\{AX-XA: X\in\mathcal{A}(n)\})={n \choose 2} - \sum_{i=1}^r { m_i \choose 2}.\]
     \item Let $W\subset\mathcal{S}(n)$ be a vector subspace containing $A$. We say that $A$ satisfies the \emph{transversality condition} if the intersection of $O(n).A$ and $W$ at $A$ is transversal, namely if
		\begin{equation}\label{eq: transversality}
			\{AX-XA:X\in\mathcal{A}(n)\}+W=\mathcal{S}(n).    
		\end{equation}
    If $A$ satisfies the \emph{transversality condition}, then, for any choice of non-vanishing eigenbasis of $A$, say $\bm{\varphi}_{1},\bm{\varphi}_{2},\ldots,\bm{\varphi}_{n}$ with $\bm{\varphi}_{k}(i)\ne 0 $ for all $i,k \in [n]$, there is an arbitrarily close matrix  $A'\in W$ that satisfies Nodal Count Condition \ref{ass} with an eigenbasis $\bm{\psi}_{1},\bm{\psi}_{2},\ldots,\bm{\psi}_{n}$, such that 
		\begin{align*}
			\mathrm{sign}(\bm{\varphi}_{k}(i))=\mathrm{sign}(\bm{\psi}_{k}(i))\ &\ \text{for all $i,k \in [n]$}, \\
			\mathrm{sign}(A'_{ij})=\mathrm{sign}(A_{ij})\ &\ \text{for all } (ij)\ \text{with} \ A_{ij}\ne0. 
		\end{align*}
 \end{enumerate}
	\end{lemma}

	\begin{proof}
	The tangent space of $O(n)$ at the identity is the Lie algebra $\mathcal{A}(n)$, so the tangent space $T_{A}(O(n).A)$ consists of all matrices of the form $\frac{d}{dt}(e^{tX}Ae^{-tX})|_{t=0}=XA-AX$ for all $X\in\mathcal{A}(n)$. Let $O_{A}\subset O(n)$ be the subgroup of $o$ for which $oAo^{T}=A$, and consider its vector space of generators $\mathcal{A}_{A}\subset \mathcal{A}(n)$, namely, $X \in \mathcal{A}(n)$ such that $e^{tX}\in O_{A} $ for all $t\in\R$. Then $\dim(\mathcal{A}_{A})=\dim(O_{A})$ and $\mathcal{A}_{A}$ is exactly the kernel of the linear map $X\mapsto XA-AX$, so $\dim(T_{A}(O(n).A))=\dim(\mathcal{A}(n))-\dim(O_{A})$. To calculate the dimension of $O_{A}$, notice that $ o\in O_{A}$, i.e., it is an orthogonal matrix that commutes with $A$, if and only if it is block diagonal with respect to the eigenspaces of $A$ and acts as an orthogonal map on each eigenspace. That is $O_{A}\cong \oplus_{i=1}^{r}O(m_{i})$ and therefore $\dim(O_{A})=\sum_{i=1}^r { m_i \choose 2}$. This proves part (1).

 For part (2), by definition, $O(n).A$ intersect $W$ transversely at $A$ (in the ambient space $\mathcal{S}(n)$) if and only if the tangent spaces $T_{A}W$ and $T_{A}(O(n).A)$ add up to $T_{A}\mathcal{S}(n)$. Using that $T_{A}W=W$ and $T_{A}\mathcal{S}(n)=\mathcal{S}(n)$, the transversality condition becomes 
 \[\{AX-AX:X\in\mathcal{A}(n)\}+W=\mathcal{S}(n).\]
		If the intersection is transversal, then for any $\epsilon>0$ there is a $\delta>0$, such that for any $\tilde{A}\in\mathcal{S}(n)$ with $\|A-\tilde{A}\|<\delta$ there is $o\in O(n)$ with $\|o-I\|<\epsilon$ for which $o\tilde{A}o^{T}\in W$. Choosing an eigenbasis gives a spectral decomposition $A=VDV^{T}$ with $D$ diagonal of eigenvalues and $V_{ik}=\bm{\varphi}_{k}(i)$. Let $D'$ be any diagonal matrix, with distinct diagonal entries, such that $\|D-D'\|<\delta$. Then  
		$\tilde{A}=VD'V^{T}$ has simple spectrum and $\|A-\tilde{A}\|=\|D-D'\|<\delta$. Letting $V'=oV$ gives
		\[A'=o\tilde{A}o^{T}=V'D'(V')^{T}\in W,\]
  so that we can set $\bm{\psi}_{k}(i)=V'_{ik}$.
		By taking $\epsilon$ and $\delta$ arbitrarily small we can make $\|A'-A\|$ and $\|V'-V\|$ arbitrarily small, which means that $|A'_{ij}-A_{ij}|$ and $|V'_{ij}-V_{ij}|$ are arbitrarily small for all $ij$, and so they have the same sign structure.
	\end{proof}

\subsection{Sparsity of the matrix inverse}

Here we provide two results regarding the sparsity structure of the inverse of a matrix $A \in \S(G)$. First, we consider the sparsity structure of the inverse of any invertible $A \in \S_0(P_{2n})$ strictly supported on $P_{2n}$, the path graph on $2n$ vertices. The following is a straightforward consequence of the recurrence formula for the entries of the inverse of a symmetric tridiagonal matrix (see \cite{usmani1994inversion} for details).

\begin{proposition}[corollary of {\cite[Theorem 1]{usmani1994inversion}}]\label{prop:tree_inverse}
If $A \in \S_0(P_{2n})$ is strictly supported on $P_{2n}$ and $\det(A) \ne 0$, then $A_{ij}^{-1} \ne 0$, $i \ge j$, if and only if $i$ is even and $j$ is odd.
\end{proposition}

In Subsection \ref{sub:treeperturb} we prove that such a matrix $A \in \S_0(P_{2n})$ always exists. Next, consider an irreducible symmetric non-negative matrix $A$. We note that $(A- \mu I)^{-1}$ is an entry-wise non-zero matrix for all but finitely many choices of $\mu \in \mathbb{R}$:

\begin{proposition}\label{prop:inverse_sparsity}
For any connected graph $G$ and non-negative matrix $A \in \S(G)$ strictly supported on $G$, there exists a finite set $\Lambda \subset \mathbb{R}$ such that $(A-\lambda I )^{-1}_{ij} \ne 0$ for all $i,j \in [n]$ and $\lambda \not \in \Lambda$.
 \end{proposition}

 \begin{proof}
Suppose $\lambda$ is not an eigenvalue of $A$. The $i,j$ entry of $(A-\lambda I )^{-1}$ is equal to zero if and only if $\mathrm{adj}(A-\lambda I)_{ij} =0$. Considering the degree $\le n-1$ polynomial $p(\lambda) = \mathrm{adj}(A-\lambda I)_{ij}$, we note that either $p(\lambda) =0$ for all $\lambda \in \mathbb{R}$, or equals zero for at most $n-1$ values of $\lambda$. Taking $\lambda> \rho(A)$, $\lambda I - A$ is an irreducible M-matrix, and so its inverse $(\lambda I - A)^{-1}$ is a positive matrix \cite[8.3.P15]{horn2012matrix}. Therefore, $p(\lambda) >0$ for $\lambda$ sufficiently large, implying that $p(\lambda) = 0$ for at most $n-1$ values of $\lambda$. Taking $\Lambda$ be the union of the eigenvalues of $A$ and the zeros of $\mathrm{adj}(A-\lambda I)_{ij}$ for all entries $i,j \in [n]$ completes the proof.
 \end{proof}

	\subsection{Nodal-Magnetic relation}\label{sub:nodalmagnetic}
	To bound the average nodal count beyond the sum of known bounds for individual eigenvectors $0\le\nu(A,\bm\varphi_{k})-(k-1)\le\beta(G)$, we exploit the topological characterization of the nodal surplus $\nu(A,\bm\varphi_{k})-(k-1)$ introduced by Berkolaiko \cite{berkolaiko2013nodal} and further developed by Colin de Verdiere \cite{colin2013magnetic}. We present this characterization in a short and simplified manner, without introducing magnetic fields and the gauge group. We recommend \cite{AlonGoresky} for a thorough introduction.  The simplified setup is as follows. Let $G$ be a simple connected graph, fix an arbitrary spanning tree of $G$, and let $((r_{\ell},s_{\ell}))_{\ell=1}^{\beta(G)} \subset E(G)$ be the $\beta(G)$ edges of $G$ that are not contained in the chosen spanning tree. Given $A\in\mathcal{S}(G)$ and $\theta=(\theta_{1},\ldots,\theta_{\beta})\in\T^{\beta}$, let $A_{\theta}$ be the Hermitian matrix supported on $G$ with
 $$(A_{\theta})_{r_{\ell},s_{\ell}}=e^{i\theta_{\ell}}A_{r_{\ell},s_{\ell}} \qquad  \text{and} \qquad  (A_{\theta})_{s_{\ell},r_{\ell}}=e^{-i\theta_{\ell}}A_{s_{\ell},r_{\ell}} \qquad \text{for all } \ell= 1,\ldots, \beta(G),$$
 and $(A_{\theta})_{ij} = A_{ij}$ otherwise.

	\begin{theorem}[\cite{berkolaiko2013nodal,colin2013magnetic} Topological characterization of the nodal surplus]\label{thm:nodal-magnetic} 
		Let $A\in\mathcal{S}(G)$ satisfy Nodal Count Condition \ref{ass}, and consider the functions $\lambda_{k}:\T^{\beta}\to\R$ with $\lambda_{k}(\theta)$ denoting the $k^{th}$ eigenvalue of $A_{\theta}$. Then $\lambda_{k}(\theta)$ is smooth around $\theta=0$, with a critical point at $\theta=0$ which is non-degenerate and has Morse index equal to $\nu(A,\bm\varphi_{k})-(k-1)$. Equivalently, $\Hess \lambda_{k}(0)$, the Hessian of $\lambda_{k}$ at $\theta=0$, is an invertible $\beta(G)\times\beta(G)$ real symmetric matrix with exactly $\nu(A,\bm\varphi_{k})-(k-1)$ negative eigenvalues.  
	\end{theorem}

This theorem, in full generality, considers (Gauge) equivalence classes of matrices. The choice of spanning tree amounts to choosing a representative $A_\theta$ from each equivalence class, which is why the above theorem is independent of the choice of spanning tree. See \cite[Subsection 2.5]{AlonGoresky} for details.

	\section{Tight Bounds for Average Nodal Count}\label{sec:avg}
	In this section, we provide a proof of Theorem \ref{thm: main bounds}, our main result. We do so in two parts. First, in Subsection \ref{sub:bound}, we prove the bounds of Theorem \ref{thm: main bounds} using Berkolaiko's theorem and a clever linear algebraic argument. Then, in Subsection \ref{sub:constructions}, we illustrate the tightness of these bounds through mathematical constructions for all choices of $n$ and $\beta$.
	
	\subsection{Proof of Theorem \ref{thm: main bounds}: Bounding nodal count}\label{sub:bound}
	Using the nodal-magnetic relation introduced in Subsection \ref{sub:nodalmagnetic}, we prove the bounds of Theorem \ref{thm: main bounds}.
	
	\begin{proof}[Proof of Theorem \ref{thm: main bounds}: Part 1]
		Following Remark \ref{rem}, we only prove the lower bound, which can be written as \[\beta(G)\le \sum_{k=1}^{n}[\nu(A,\bm\varphi_{k})-(k-1)].\]  
		Let $H_{k}=\Hess \lambda_{k}(0)$, as described in Theorem \ref{thm:nodal-magnetic}, so that $\det(H_{k})\ne 0$ and $H_k$ has exactly $\nu(A,\bm\varphi_{k})-(k-1)$ negative eigenvalues, denoted by $\mathrm{ind}(H_{k})=\nu(\bm\varphi_{k})-(k-1)$. We aim to show that 
		\[\beta(G)\le \sum_{k=1}^{n}\mathrm{ind}(H_{k}).\]
		Suppose, for the sake of contradiction, that $\sum_{k=1}^{n}\mathrm{ind}(H_{k})=m<\beta$. Let $V_{k}\subset\R^{\beta}$ be the $\mathrm{ind}(H_{k})$ dimensional vector space on which $H_{k}$ is negative-definite. Since $H_{k}$ is invertible, it is positive-definite on $V_{k}^\perp$, namely $\langle H_{k} u,  u \rangle>0$ for all non-zero $u \in V_{k}^\perp$. The vector space $V=V_{1}+V_{2}+\ldots+V_{n}$ has dimension at most $m<\beta$, so $\bigcap_{k=1}^{n}V_{k}^\perp$ has positive dimension, and hence contains at least one non-zero vector $u$. For this vector, $\langle H_{k} u, u\rangle>0$ for all $k \in [n]$, which means that $\langle \left(\sum_{k=1}^{n}H_{k}\right) u, u\rangle>0$. However, by construction, $\mathrm{trace}(A_{\theta})$ is independent of $\theta \in\T^{\beta}$, so its Hessian is the zero matrix, $\sum_{k=1}^{n}H_{k}=\sum_{k=1}^{n}\Hess \lambda_{k}(0)=\Hess\left(\mathrm{trace}(A_{\theta})\right)=0$, and therefore $\langle \left(\sum_{k=1}^{n}H_{k}\right) u, u\rangle=0$, providing the needed contradiction.
	\end{proof}
	
	\subsection{Proof of Theorem \ref{thm: main bounds}: Construction of extremizers 
		$G_{n,\beta}$ }\label{sub:constructions}
	\begin{definition}    
		Given $n \in \mathbb{N}$ and $0 < \beta \le {n-1 \choose 2}$, 
		let $G_{n,\beta} = ([n],E)$, where
		$$E = \{(1,i) \, | \, \ell+1<i \le m\} \cup \{(i,j) \,| \, 1<i<j\le m\} \cup \{(i,i+1)\, | \, m \le  i < n\},$$
		and $m$ and $\ell$ are such that ${m-2 \choose 2}<\beta\le {m-1 \choose 2}$ and $\ell={m-1 \choose 2}-\beta$ (notice that $\ell\le m-2$). This graph is a clique, with some edges removed, attached to a path (see Figure \ref{fig:minimizers}).
	\end{definition}
Here, and in what follows, we use the standard convention that ${n \choose k} = 0$ for all $n<k$. We note that $G_{n,\beta}$ only exists for $n >2$, as the unique connected graphs on one and two vertices are both trees. In this subsection, we construct a matrix $A\in\mathcal{S}(G_{n,\beta})$, for every choice of $n \in \mathbb{N}$ and $0 \le \beta \le {n-1 \choose 2}$, that satisfies (NCC) and has $\sum_{k=1}^{n}\nu(A,\bm\varphi_{k})={n \choose 2}+\beta$. This construction illustrates the tightness of Theorem \ref{thm: main bounds}, thus completing its proof, as the case $\beta(G) = 0$ is simply Fiedler's matrix tree theorem \cite{fiedler1975eigenvectors}. We first treat the regime ${n-2 \choose 2}< \beta \le {n-1 \choose 2}$ (i.e., $m = n$). 
	\begin{proposition}\label{prop:dense construction}
		Given $ n\in\N $, $n >2$, and ${n-2 \choose 2}< \beta \le {n-1 \choose 2}$, the matrix $ A \in\S(G_{n,\beta})$, defined below, has the following properties:
		\begin{enumerate}
			\item it has a choice of nowhere-vanishing eigenbasis $\bm{\varphi}_{1},\ldots,\bm{\varphi}_{n}$ with total nodal count
   \[\sum_{k=1}^{n}\nu(A,\bm\varphi_{k})={n \choose 2}+\beta,\]
			\item it satisfies the transversality condition of Lemma \ref{lem: transversality} with respect to $W=\S(G_{n,\beta})$.
		\end{enumerate}
	Therefore, there exists an $A' \in\S(G_{n,\beta})$ arbitrarily close $A$ that satisfies Nodal Count Condition \ref{ass} and has the same total nodal count $ {n \choose 2}+\beta $. The matrix $A$ is given by 
	\[ A = \begin{blockarray}{cccc}
		{\scriptstyle 1} & {\scriptstyle \ell} & {\scriptstyle n-\ell-1} & \\
		\begin{block}{[ccc]c}
			(n-1)^3 - \ell {n-1 \choose 2} & 0 & -(n-1){n-1 \choose 2} \mathbf{1}^T &{ \scriptstyle 1} \\ 
			0 & (n-1) {n \choose 2} I -(n-\ell-1)\mathbf{1}\mathbf{1}^T &  -\big[{n \choose 2} - \ell \big]\mathbf{1}\mathbf{1}^T & {\scriptstyle \ell} \\
			-(n-1){n-1 \choose 2} \mathbf{1} & -\big[{n \choose 2} - \ell \big]\mathbf{1}\mathbf{1}^T & {n \choose 2}^2 I - \big[(n-1)^2 - \ell \big]\mathbf{1}\mathbf{1}^T & {\scriptstyle n-\ell-1} \\
		\end{block}
	\end{blockarray}\]
where  $\ell = { n-1 \choose 2} - \beta$, $I$ denotes the identity matrix, $0$ denotes the all-zeros vector, and $\mathbf{1}$ denotes the all-ones vector, with the dimension of each clear from context. If $\ell = 0$, then $A$ consists only of the first and third rows and columns of the above block matrix.
	\end{proposition}
	\begin{proof}
		By inspection, $A$ admits the orthonormal eigendecomposition
		\begin{align*}
			\lambda_1 &= \ell {n \choose 2} ,  & \bm\varphi_1 &= \frac{1}{n} \begin{bmatrix} n-2 \\ 2 \mathbf{1}_{n-1} \end{bmatrix}, \\
			\lambda_k &= \begin{cases} (n-1){n \choose 2} & 1< k \le \ell+1 \\  {n \choose 2}^2 & \ell+1<k \le n \end{cases},  & \bm\varphi_k &= {n \choose 2}^{-1} \begin{bmatrix} n-1 \\ \mathbf{1}_{n-1} -{n \choose 2} \bm e_{k-1} \end{bmatrix},
		\end{align*}
		where $\bm{e}_k$ denotes the $k^{th}$ standard basis vector. The total nodal count of the eigenbasis $\{\bm\varphi_1,\ldots,\bm\varphi_n\}$ is 
		\begin{align*}
			\nu(A,\bm\varphi_1) + \sum_{k = 2}^{\ell+1} \nu(A,\bm\varphi_k) + \sum_{k = \ell+2}^{n} \nu(A,\bm\varphi_k) &= 0 + \ell (n-2) + (n- \ell -1 )(n-1) \\  &= (n-1)^2 - \ell \\ &= {n \choose 2} + \beta(G).
		\end{align*}
		This proves property (1). To show property (2), note that the set of non-edges of $ G $ consists of all $(1j)$ with $j\in J = \{2,\ldots,\ell+1\}$. Transversality condition \eqref{eq: transversality} for $W=\mathcal{S}(G_{n,\beta})$ holds if, for any matrix $C\in\mathcal{S}(n)$, there is a matrix  $X\in\mathcal{A}(n)$ such that $(XA -AX)_{1j}=C_{1j}$ for all $j\in J$. If $\ell = 0$, then $J = \emptyset$ and we are done. Now, suppose $J$ is non-empty and consider $X\in\mathcal{A}(n)$ with entries $X_{ij} = 0$, $i<j$, for $(i,j) \not \in \{1\} \times J$. In this case, for $i \in J$, we have 
		\begin{align*}
			(XA-AX)_{1i} &= X_{1i} (A_{ii}-A_{11})+ \sum_{ k\in J, k \ne i} X_{1k} A_{ki}\\
			&= -(n-\ell-1)\bigg[{n-1 \choose 2} X_{1i}+\sum_{k \in J} X_{1k} \bigg],
		\end{align*}
		implying that $(XA-AX)_{1i}$, $i\in J$, are linearly independent (linear) functions in $(X_{1k})_{k\in J}$, so for any choice of $(C_{1,j})_{j\in J}$ there is an $X$ as above for which $(XA-AX)_{1i}=C_{1i}$ for all $i\in J$. Therefore, the transversality condition holds, and by Lemma \ref{lem: transversality} there is $A'\in\mathcal{S}(G_{n,\beta})$ satisfying (NCC) with the same nodal count as $A$.  
	\end{proof}

	This treats the regime ${n-2 \choose 2}< \beta \le {n-1 \choose 2}$. To produce constructions $A \in \mathcal{S}(G_{n,\beta})$ for $0<\beta \le {n-2 \choose 2}$, we prove an interpolation lemma that analyzes the effect on nodal count of adding an edge of small weight between two graphs, one of which contains a vertex of full degree.
	
	\begin{lemma}\label{lm:interpolation}
		Let $G_{1},G_{2}$ be two connected graphs with $n_1$ and $n_2$ vertices respectively, and suppose that vertex $n_{1}$ of $G_{1}$ is connected to all other vertices of $G_{1}$. 
		Let $G$ be the graph on $ n=n_{1}+n_{2} $ vertices obtained by taking $G_1$ and $G_2$ and adding an edge between vertex $ n_{1} $ of $ G_{1} $ and vertex $ 1 $ of $G_{2}$, so that $\beta(G)=\beta(G_{1})+\beta(G_{2})$. Given any two matrices   
		 $A_1\in\mathcal{S}(G_{1})$ and $A_2\in\mathcal{S}(G_{2})$ that satisfy Nodal Count Condition \ref{ass}, have only non-positive entries, and have total nodal count $\nu_1$ and $\nu_2$ respectively, there is a sufficiently large spectral shift $s\gg 1$ such that, for all sufficiently small $\epsilon >0$, the matrix
  \[ A_{\epsilon} = \begin{blockarray}{ccc}
		{\scriptstyle n_1} & {\scriptstyle n_2} &  \\ 
  \begin{block}{[cc]c} 
  A_1 - s I & -\epsilon \, \bm{e}_{n_1} \bm{e}_1^T & {\scriptstyle n_1} \\ -\epsilon \, \bm{e}_1 \bm{e}_{n_1}^T & A_2 & {\scriptstyle n_2} \\ 
  \end{block} 
  \end{blockarray} \in \mathcal{S}(G) \]
satisfies Nodal Count Condition \ref{ass} and has total nodal count equal to $\nu_1 + \nu_2 + n_1 n_2$.
	\end{lemma}
	
	\begin{proof}
		Let $A=(A_{1}-sI)\oplus A_{2}$ for a fixed sufficiently large $ s $, (to be determined later), and let $B= -\bm{e}_{n_1} \bm{e}_{n_{1}+1}^T-\bm{e}_{n_{1}+1} \bm{e}_{n_{1}}^T$ so that $A_{\epsilon}=A+\epsilon B$ as in the setting of Lemma \ref{lm: perturbation simple}. Note that by taking  $ s $ sufficiently large, the spectra of $ A_{1}-sI $ and $ A_{2} $ are disjoint, so $ A $ has a simple spectrum. Lemma \ref{lm: perturbation general} provides a small interval $ (0,b) $ so that, for all $ \epsilon\in(0,b) $, the eigenvalues of $A_{\epsilon}$ are simple and the support and signs of the eigenvectors are independent of $ \epsilon $. For $ \epsilon\in(0,b) $, an eigenpair of $A_{\epsilon}$ has the form  $\lambda_{\epsilon}=\lambda^{(0)}+\epsilon \lambda^{(1)} + O(\epsilon^2) $ and $ \bm\varphi_{\epsilon}=\bm\varphi^{(0)}+\epsilon \bm\varphi^{(1)}+O(\epsilon^{2})$. We break our analysis into two cases, depending on whether $\lambda^{(0)}$ is an eigenvalue of $A_1 - sI$ or $A_2$.

	First, we consider the case when $\lambda^{(0)}$ is an eigenvalue of $A_{1}-sI$ with non-vanishing eigenvector $\bm{u}\in\R^{n_{1}}$, so that $ \bm\varphi^{(0)}=(\bm{u},0)^T$. Then, Lemma \ref{lm: perturbation simple} and the block structure of $ A $ and $ B $ implies that $\lambda^{(1)} = \langle B \bm{\varphi}^{(0)}, \bm{\varphi}^{(0)} \rangle = 0$ and 
		\[\bm\varphi^{(1)}=-(A-\lambda^{(0)}I)^{+}B\bm\varphi^{(0)}=\begin{bmatrix}0 \\ \bm{u}(n_{1})(A_{2}-\lambda^{(0)}I)^{-1}\bm{e}_{1} \end{bmatrix}.\]
		By making the shift $s$ sufficiently large, we ensure that $ \lambda^{(0)}<0$ and $ |\lambda^{(0)}|$ is strictly larger than the spectral radius of $A_{2}$, so $ A_{2}-\lambda^{(0)}I $ is an $ M $-matrix. Because $A_2$ is irreducible ($G_2$ is connected), $(A_{2}-\lambda^{(0)}I)^{-1}$ is an entrywise positive matrix and so $ (A_{2}-\lambda^{(0)}I)^{-1}\bm{e}_{1} $ is a non-vanishing positive vector \cite[8.3.P15]{horn2012matrix}. 
		We conclude, using Lemma \ref{lm: perturbation general}, that $ \bm\varphi_{\epsilon} $ is non-vanishing, with constant sign on the vertices $n_{1}\le j\le n_{1}+n_{2}$ and with the same sign pattern as $\bm{u}$ on the vertices $1\le j\le n_{1}$. The nodal count is therefore
		\[\nu(A_{\epsilon},\bm\varphi_{\epsilon})=\nu(A_1,\bm{u}) .\]

	Next, we consider the case when $ \lambda^{(0)} $ is an eigenvalue of $A_{2}$ with non-vanishing eigenvector $\bm{u} \in\R^{n_{2}}$, so that now $ \bm\varphi^{(0)}=(0,\bm{u})^T$. The same argument implies that
		\[\bm\varphi^{(1)}= \begin{bmatrix} \bm{w} \\ 0 \end{bmatrix} \quad\text{with}\quad \bm{w}=\bm{u}(1)
\left[(A_{1}-sI)-\lambda^{(0)}I\right]^{-1}\bm{e}_{n_1}.\]  
		For notational convenience, let $ \delta=(s+\lambda^{(0)})^{-1} $. Let $s$ be sufficiently large so that $\delta>0$ and the spectral radius of $ \delta A_{1} $ is less than one. By expanding the inverse as a power series 
  \[(I- \delta A_{1})^{-1}=\sum_{m=0}^{\infty}\delta^{m}A_1^{m}=I+\delta A_1+O(\delta^{2}),\]
  we have
		\[\bm{w}=- \delta \bm{u}(1)(I-\delta A_{1})^{-1}\bm{e}_{n_1}=- \delta \bm{u}(1)\left[\bm{e}_{n_1}+\delta A_{1}\bm{e}_{n_1}+O(\delta^{2})\right].\]
		Because vertex $n_{1}$ of $ G_1$ is connected to all other vertices of $G_1$, $(A_{1})_{j,n_{1}}<0$ for all $j<n_{1}$. This means that, for $\delta>0$ sufficiently small, $\bm{w}$ is non-vanishing with $\mathrm{sign}(\bm{w}(n_1)) = - \mathrm{sign}(\bm{u}(1))$ and $\mathrm{sign}(\bm{w}(j))=\mathrm{sign}(\bm{u}(1))=-\mathrm{sign}(\bm{w}(n_1))$ for all $j<n_{1}$. Since the sign structure of $ \bm\varphi_{\epsilon} $ is equal to the sign structure of $\bm{w}$ on $G_1$ and $\bm{u}$ on $G_2$, then its nodal count is that of $\bm{u}$ plus a sign change on the edge $(n_{1},n_{1}+1)$ and sign changes on the $n_1 -1$ edges $ (j,n_{1}) $, $ j<n_{1} $. Namely,  \[\nu(A_{\epsilon},\bm\varphi_{\epsilon})=\nu(A_2,\bm{u})+n_{1} .\]

	Summing the nodal count in each of the above two cases, we conclude that the total nodal count of $ A_{\epsilon} $ is $\nu_{1}+\nu_2 + n_1 n_2$.
	\end{proof}

We are now prepared to complete the proof of Theorem \ref{thm: main bounds}. In particular, we are prepared to prove that, for any $n\in\N$, $n>2$, and $ 0< \beta\le{n-1 \choose 2}$, there is a matrix $ A\in\mathcal{S}(G_{n,\beta}) $ that satisfies (NCC) and has total nodal count $ {n \choose 2}+\beta$ (recall, the case $\beta = 0$ is Fielder's matrix tree theorem \cite{fiedler1975eigenvectors}). 
	
\begin{proof}[Proof of Theorem \ref{thm: main bounds}: Part 2]
	Let $ m $ such that $ 
	{m-2 \choose 2}< \beta\le {m-1 \choose 2} $, so that $ G_{n,\beta} $ is obtained by connecting the graph $ G_{m,\beta} $ and an $(n-m)$-vertex path graph. Proposition \ref{prop:dense construction} produces a matrix $ A_{1}\in\mathcal{S}(G_{m,\beta}) $ that satisfies (NCC), has non-positive off-diagonal entries (and can be shifted to obtain negative diagonal entries), and has total nodal count $\nu_{1}={m \choose 2}+\beta$. For the path graph $ G_{2} $ on $n-m$ vertices, let $ A_{2}\in \mathcal{S}(G_{2})$ be any non-positive matrix with simple spectrum (such a matrix exists by genericity). By Fiedler's matrix tree theorem \cite{fiedler1975eigenvectors}, $A_{2}$ must satisfy (NCC) and its total nodal count must equal $ \nu_{2}= {n-m \choose 2} $. We may now apply Lemma \ref{lm:interpolation} to $G_1:=G_{m,\beta}$ and $G_2$ to obtain the matrix $ A_{\epsilon}\in\mathcal{S}(G_{n,\beta})$ which satisfies (NCC) and whose total nodal count is  
	\[\nu_{1}+\nu_{2}+m(n-m)= {m \choose 2} + \beta + {n-m \choose 2} + m(n-m) = {n \choose 2} +\beta.\]
	
\end{proof}

\subsection{Nodal count for $\S_0(G)$}

	In the case of $\mathcal{S}_0(G)$, we note that the bounds of Theorem \ref{thm: main bounds} can still be tight. Here, we consider dense constructions in $\mathcal{S}_0(K_n)$ that achieve the bounds of Theorem \ref{thm: main bounds}. We consider sparse zero-diagonal constructions, as well as alternate constructions for $\mathcal{S}(G)$ for the regime $0<\beta\le n-2$, in the following subsection.

 \begin{proposition}
 Let $n \in \N$, $n >2$, and $K_n$ be the complete graph on $n$ vertices. For $\epsilon>0$ sufficiently small, the matrix $ A'=A_\epsilon -I \in\S_0(K_n)$, with $A_\epsilon$ defined below, has the following properties:
		\begin{enumerate}
			\item it has a choice of nowhere-vanishing eigenbasis $\bm{\varphi}_{1},\ldots,\bm{\varphi}_{n}$ with total nodal count \[ \sum_{k=1}^{n}\nu(A,\bm\varphi_{k})={n \choose 2}+\beta(K_n),\]
			\item it satisfies the transversality condition of Lemma \ref{lem: transversality} with respect to $W=\S_0(K_n)$.
		\end{enumerate}
	Therefore, there exists an $ A'' \in\S_0(K_n)$ arbitrarily close to $A'$ that satisfies Nodal Count Condition \ref{ass} and has the same total nodal count $ {n \choose 2}+\beta(K_n) $. The matrix $A_\epsilon$ is given by 
 	\[ A_\epsilon = - \; \begin{blockarray}{cccc}
 {\scriptstyle 1} &  {\scriptstyle 1} &  {\scriptstyle n-2} & \\
 \begin{block}{[ccc]c} \epsilon^2 & 1-\epsilon^2 & a \, \epsilon \,\bm{1}^T & {\scriptstyle 1} \\
		1-\epsilon^2 & \epsilon^2 & b \, \epsilon \,\bm{1}^T & {\scriptstyle 1}\\
		a \, \epsilon  \, \bm{1} & b \, \epsilon \,\bm{1} & \epsilon^2 \bm{1} \bm{1}^T & {\scriptstyle n-2} \\ \end{block}
  \end{blockarray}\]
where $a = (1+\frac{9\epsilon^2}{4+\epsilon^2})\frac{1}{2\sqrt{2}}$, $b = (1-\frac{9\epsilon^2}{8+2 \epsilon^2})\sqrt{2}$, and $\mathbf{1}\in\R^{n-2}$ denotes the all-ones vector.
	\end{proposition}

 \begin{proof}
Notice that $A_{\epsilon}=A'+I$, so $A_{\epsilon}$ and $A'$ share the same eigenvectors and same nodal count, $\sum_{k=1}^{n}\nu(A_{\epsilon},\bm\varphi_{k})=\sum_{k=1}^{n}\nu(A',\bm\varphi_{k})$. The matrix $A_{\epsilon}$ is rank two, with the vector \[\bm{x} = \begin{bmatrix} (8-\epsilon^2)(n-2) \epsilon \\  (2-\epsilon^2)(n-2) \epsilon \\ -\sqrt{2}(4+\epsilon^2)  \, \bm{1}_{n-2} \end{bmatrix}\] 
in its nullspace, as well as any vector $(0,0,\bm{y}^T)^T$ with $\bm{y}^T \bm{1} = 0$. For $\epsilon$ sufficiently small, the two eigenvectors not in the nullspace are of the form $ \bm{\varphi}_1 = \frac{1}{\sqrt{2}}(1,1,\bm{0}_{n-2})^T + O(\epsilon)$ and $\bm{\varphi}_n = \frac{1}{\sqrt{2}}(1,-1, \bm{0}_{n-2}^T) + O(\epsilon)$ by \ref{lm: katto-rellich}. By orthogonality to the nullspace, both are constant on the indices $\{3,\ldots, n\}$. For $\epsilon$ sufficiently small,
\[ \mathrm{null}\left( \begin{bmatrix} \lambda + \epsilon^2 & 1-\epsilon^2 \\ 1- \epsilon^2 & \lambda + \epsilon^2 \\ a \, \epsilon  & b \, \epsilon \end{bmatrix} \right) = 0 \qquad \text{for all } \lambda \in \mathbb{R},\]
implying that both $\bm{\varphi}_1$ and $\bm{\varphi}_n$ are also non-vanishing on the indices $\{3,\ldots, n\}$. Therefore, for $\epsilon$ sufficiently small, $\bm{\varphi}_1$ and $\bm{\varphi}_n$ are non-vanishing and have nodal counts $\nu(A_{\epsilon},\bm{\varphi}_1) = 0$ (by Perron-Frobenius) and $\nu(A_{\epsilon},\bm{\varphi}_n) =n-1$.

Now, consider the non-vanishing orthonormal basis
\[ \bm{\varphi_{k}} = \bm{x} + \frac{\|\bm{x}\|}{\sqrt{n-2}} \begin{bmatrix} 0 \\ 0 \\ \bm{1} - (n-2) \bm{e}_{k-1}\end{bmatrix}, \qquad k = 2,\ldots,n-1,\]
for the nullspace of $A_\epsilon$. For $k \in \{2,\ldots,n-1\}$, we have $\bm{\varphi}_k(1)>0$, $\bm{\varphi}_k(2)>0$, $\bm{\varphi}_k(k+1)<0$, and
\[ \bm{\varphi}_k(\ell) = -\sqrt{2}(4+\epsilon^2) + \sqrt{2\,(4+\epsilon^2)^2 + (n-2)[(8-\epsilon^2)^2 + (2-\epsilon^2)^2] } >0 \]
for all $\ell \ne 1,2,k+1$. Therefore, $\nu(A_{\epsilon},\bm{\varphi}_k) = n-1$ for all $k = 2,\ldots n-1$, and so 
\[ \sum_{k=1}^{n}\nu(A,\bm\varphi_{k})= (n-1)^2 = {n \choose 2} + {n-1 \choose 2} = {n \choose 2} + \beta(K_n)\] 
 for this eigenbasis of $A_\epsilon$. This proves property (1). 

 Next, we consider property (2). Because $G = K_n$ is the complete graph, the orthogonal complement of $\mathcal{S}_{0}(K_n)$ is the space of diagonal matrices with trace zero. Transversality condition \ref{eq: transversality} holds if, for any $\bm{v}\in\R^{n}$ with $\langle \bm{v},\textbf{1}\rangle=0$, there exists an $X\in\mathcal{A}(n)$ such that $(A'X-XA')_{jj}= \bm{v}(j)$ for all $j$. We calculate
	\[\frac{1}{2}(A'X-XA')_{jj}=(A'X)_{jj}=\begin{cases}
		(\epsilon^2 -1)X_{12} - a \, \epsilon \sum_{i=3}^{n}X_{1i}\ & \ \text{for } j=1\\
		(\epsilon^2 -1)X_{21} - b \, \epsilon \sum_{i=3}^{n}X_{2i}\ & \ \text{for } j=2\\
		-a \, \epsilon X_{j1} - b \, \epsilon X_{j2} - \epsilon^2 \sum_{i=3}^{n}X_{ji}\ & \ \text{for } j>2
	\end{cases}.\]
Let $X^{(i,j)} \in \mathcal{A}(n)$, $i < j$, be the anti-symmetric matrix with $X^{(i,j)}_{ij} = -X_{ji}^{(i,j)} =1$ and equal to zero elsewhere. We note that, for $j>2$,
 \[(A'X^{(1,j)})_{ii}= \begin{cases}
		-a \, \epsilon \ & \ \text{for } i=1\\
		\phantom{-}a \, \epsilon \ & \ \text{for } i=j\\
		\phantom{-}0 \ & \ \text{otherwise}
	\end{cases} \qquad \text{and} \qquad (A'X^{(2,j)})_{ii}= \begin{cases}
		-b \, \epsilon \ & \ \text{for } i=2\\
		\phantom{-}b \, \epsilon \ & \ \text{for } i=j\\
		\phantom{-}0 \ & \ \text{otherwise}
	\end{cases} .\]
  Any $\bm{v}\in\R^{n}$ with $\langle \bm{v},\textbf{1}\rangle=0$ is in the span of the vectors $\{\bm{e}_1 -\bm{e}_j\}_{j=3}^n \cup \{\bm{e}_2 - \bm{e}_j \}_{j=3}^n$, completing the proof.
 \end{proof}

	Finally, we complement the above dense constructions with sparse ones, using perturbations of tridiagonal matrices. In particular, we show that, even though trees produce matrices with strongly controlled nodal edge count, we can still produce perturbations that lead to fairly extreme behavior of the nodal surplus. We do so by first producing an alternate tight construction for Theorem \ref{thm: main bounds} in the regime $0 < \beta \le n-2$, and then adjusting this construction to obtain a constant diagonal construction with nodal surplus $2 \beta$ for a similar regime.
	
	\subsection{Path-based constructions for small $\beta$}\label{sub:treeperturb}
\begin{figure}
      \centering
	   \begin{subfigure}[Extremizer $H_{n,\beta}$ with $n=8$ and $\beta=3$. ]  
		{\includegraphics[width=10cm]{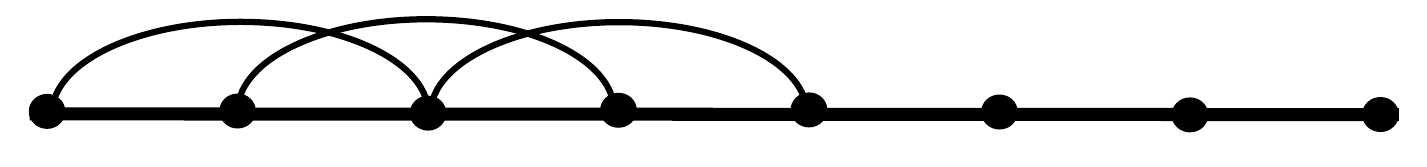} \label{fig:minimizers2a}}

	   \end{subfigure}
	\vfill
	     \begin{subfigure}[Extremizer $H_{2n,\beta}^{(0)}$ with $n=8$ and $\beta=3$.]
		{ \includegraphics[width=10cm]{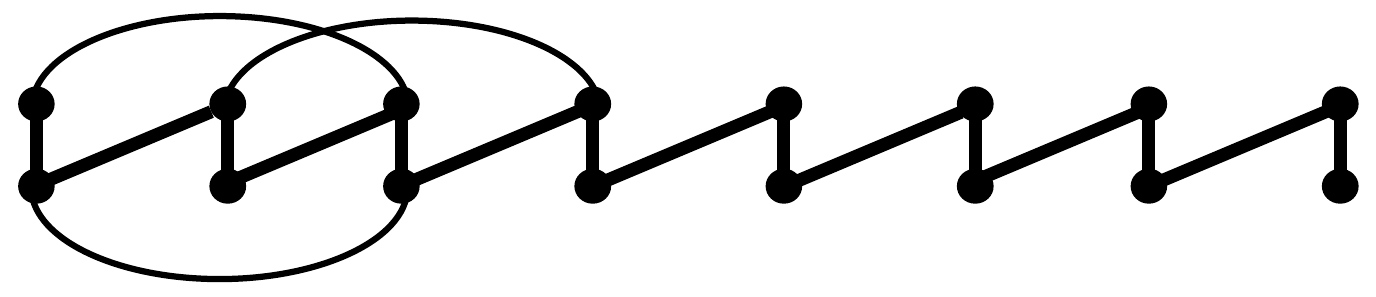} \label{fig:minimizers2b}}
	      \end{subfigure}
	\caption{Example of extremizers from path constructions.}
	\label{fig:minimizers2}
\end{figure}

In this subsection, we provide another class of extremizers based on a different approach. We construct a tridiagonal matrix (a matrix supported on the path graph $P_{n}$) whose eigenvectors have a special sign structure: $\bm\varphi_{k}(i)\bm\varphi_{k}(i+2)$ is negative if and only if $i=k$. The extremizer graphs $H_{n,\beta}$ are obtained by adding $\beta$ edges of the form $(i,i+2)$. We also provide examples of zero-diagonal matrices for which the bounds of Theorem \ref{thm: main bounds} are nearly tight. We do so by constructing zero-diagonal matrices on $P_{2n}$ that satisfy (NCC) and have a similar eigenvector structure. The existence of matrices on $P_{2n}$ that satisfy (NCC) is also a key ingredient in the proof of Theorem \ref{thm: NCC characterization intro} in Section \ref{sec:ncc} (see Remark \ref{rem:NCC on P2n} and the proof of Lemma \ref{lm:cycle} for details).  
\begin{definition}
\label{def:pathExt}
Given $n \in \mathbb{N}$ and $0 \le \beta \le n-2$, let $H_{n,\beta} = ([n],E)$, where
\begin{align*}
E &= \{(i,i+1) \, | \, 1 \le i < n \} \cup  \{(i,i+2)\, | \, 1 \le  i  \le \beta \}.
\end{align*}
This graph is a path with $\beta$ additional edges added, all between vertices at path distance two (see Figure \ref{fig:minimizers2a}).
\end{definition}

\begin{definition}    
Given $n \in \mathbb{N}$ and $0 \le \beta \le 2(n-2)$, let $H^{(0)}_{2n,\beta} = ([2n],E_0)$, where
\begin{align*}
E_0 &= \{(i,i+1) \, | \, 1 \le i < 2n \} \cup  \{(i,i+4)\, | \, 1 \le  i  \le \beta \}.
\end{align*}
This graph is a path with $\beta$ additional edges added, all between vertices at path distance four (see Figure \ref{fig:minimizers2b}).
\end{definition}
In this subsection and the next, it is useful to count the number of sign changes $A_{ij}\bm\varphi_{k}(i)\bm\varphi_{k}(j)>0$ per edge (over all eigenvectors) rather than per eigenvector (over all edges). We introduce the following notation.

\begin{definition}\label{def:edge NC}    
Let $(ij)$ be an edge of a graph $G$ and $A\in\mathcal{S}(G)$ with simple eigenvalues and eigenbasis $\bm\varphi_{1},\bm\varphi_{2},\ldots,\bm\varphi_{n}$. The contribution of $(ij)$ to the total nodal count is denoted by 
\begin{equation}\label{eq:ENC}
\nu(A,(ij)):=\left|\{k\in[n]: A_{ij}\bm\varphi_{k}(i)\bm\varphi_{k}(j)>0\}\right|.
\end{equation}
\end{definition}  

\begin{proposition}\label{prop:path}
Given $n \in \N$, $0 \le \beta \le n-2$, $\epsilon>0$ sufficiently small, and $\delta>0$ sufficiently small relative to $\epsilon$, let
\[M = D(\bm{u}) - \epsilon A - \delta B \in \S(H_{n,\beta}),\]
where $\bm{u} = (1,2,\dots, n)^T$, $D(\bm{u})$ is the diagonal matrix with $\bm{u}$ on the diagonal, $A$ is the adjacency matrix of the path $P_n$ on $n$ vertices, and $B$ is the adjacency matrix of $H_{n,\beta}$. The matrix $M$ satisfies Nodal Count Condition \ref{ass}, has eigenvectors $\bm{\varphi}_1,\ldots, \bm{\varphi}_n$ with sign structure
\[ \mathrm{sign}\big(\bm{\varphi}_k(\ell)\big) = \begin{cases} (-1)^{k-\ell} & \text{ for } \ell<k \\ \, +1 & 
			\text{ for } \ell\ge k  \end{cases},\]
and therefore has total nodal count
\[ {n \choose 2} + \beta.\]
\end{proposition}

\begin{proof} We consider $M=M_{\epsilon}-\delta B$ as a $\delta>0$ perturbation of the matrix $M_\epsilon := D(\bm{u}) - \epsilon A$ which is strictly supported on the path $P_{n}$. We first show that, for sufficiently small $\epsilon$, the matrix $M_{\epsilon}$ has simple eigenvalues and non-vanishing eigenvectors $\bm{\psi}_1,\ldots, \bm{\psi}_n$ with sign structure
$$ \mathrm{sign}\big(\bm{\psi}_k(\ell)\big) = \begin{cases} (-1)^{k-\ell} & \text{ for } \ell<k \\ \, +1 & 
			\text{ for } \ell\ge k  \end{cases}.$$
The matrix $M_{\epsilon}$ is an $O(\epsilon)$ perturbation of $D(\bm u)$, which has distinct diagonal entries, and so $M_{\epsilon}$ has simple eigenvalues for all $\epsilon\in (-T,T)$ for some $T>0$ (specified in Lemma \ref{lm: perturbation simple}). Let $\lambda_\epsilon = \sum_{i=0}^\infty \epsilon^i \lambda^{(i)}$ be the $k^{th}$ smallest eigenvalue of $M_\epsilon$ and $\bm\psi_{k}=\bm\psi_{\epsilon}=\sum_{i=0}^\infty \epsilon^i \bm{\psi}^{(i)}$ be the corresponding eigenvector, with $\lambda^{(0)} = k$ and $\bm\psi^{(0)}=\bm e_{k}$. By Lemma \ref{lm: perturbation general}, it suffices to analyze the value of $\bm{\psi}^{(j)}(\ell)$ for the first $j$ for which $\bm{\psi}^{(j)}(\ell) \ne 0$. Since $(D(\bm u)-kI)^{+}$ is diagonal with $(D(\bm u)-kI)^{+}_{\ell \ell} = \frac{1}{\ell-k}$ for $\ell \ne k$, Equation \eqref{eq: pert vj full} implies that
\[\bm{\psi}^{(j)}(\ell)=\frac{1}{\ell-k}\left[(A\bm{\psi}^{(j-1)})(\ell)+\sum_{i=1}^{j}\lambda^{(i)}\bm{\psi}^{(j-i)}(\ell)\right] \qquad \text{for all } \ell \ne k.\]
We claim that the support of $\bm{\psi}^{(j)}$ is contained in the interval $[k-j,k+j]$. We proceed by induction on $j$. When $j=0$ we have $\bm \psi^{(0)}=\bm e_{k}$. Now suppose the desired claim holds for $\bm{\psi}^{(0)}, \ldots, \bm{\psi}^{(j-1)}$. Then, for $\ell \notin [k-j+1,k+j-1]$, the sum on the right-hand side of the above formula vanishes, giving 
\begin{align*}
  \bm{\psi}^{(j)}(\ell)=
 \frac{1}{\ell-k}\bm{\psi}^{(j-1)}(\ell+1)\  &\quad \text{for}\ 1 \le \ell\le k-j,\\
 \bm{\psi}^{(j)}(\ell)=
 \frac{1}{\ell-k}\bm{\psi}^{(j-1)}(\ell-1)\  &\quad \text{for}\ k+j \le \ell \le n,
\end{align*}
and so $\bm{\psi}^{(j)}$ is supported inside $[k-j,k+j]$. In addition, if $k-j \in [n]$, then
\[\mathrm{sign}\big(\bm{\psi}^{(j)}(k-j)\big) =-\mathrm{sign}\big(\bm{\psi}^{(j-1)}(k-(j-1))\big)=\ldots= (-1)^{j}\mathrm{sign}\big(\bm{\psi}^{(0)}(k)\big)=(-1)^{j},\]
and, similarly, if $k+j \in [n]$, then
\[\mathrm{sign}\big(\bm{\psi}^{(j)}(k+j)\big) =\mathrm{sign}\big(\bm{\psi}^{(j-1)}(k+(j-1))\big)=\ldots=\mathrm{sign}\big(\bm{\psi}^{(0)}(k)\big)=1.\]
Therefore, $M_{\epsilon}$ has simple spectrum and non-vanishing eigenvectors with the desired sign structure. 

We now consider $M=M_{\epsilon}-\delta B$ as a $\delta$ perturbation of $M_{\epsilon}$. For sufficiently small $\delta$ (with respect to the eigenvalues of $M_{\epsilon}$ and $B$), $M$ has simple eigenvalues by Lemma \ref{lm: perturbation simple}, and non-vanishing eigenvectors $\bm{\varphi}_1,\ldots, \bm{\varphi}_n$ with $\mathrm{sign}(\bm \varphi _{k}(\ell))=\mathrm{sign}(\bm\psi _{k}(\ell))$ for all $k,\ell\in[n]$ by Lemma \ref{lm: perturbation general}. Because $M$ and $M_{\epsilon}$ have the same sign pattern on $P_{n}$, the difference in the nodal count is only due to the contribution of the $\beta$ additional $(i,i+2)$ edges. The sign structure of the eigenvectors $\bm{\varphi}_k$ are such that $\bm\varphi_{k}(i)\bm\varphi_{k}(i+2)<0$ if and only if $k=i+1$, implying that $\nu(M,(i,i+2)) = 1$ (see Equation \eqref{eq:ENC} for definition) for all $(i,i+2) \in E(H_{n,\beta})$. Therefore,
\begin{align*}
    \sum_{k=1}^{n}\nu(M,\bm\varphi_{k})= & \sum_{k=1}^{n}\nu(M_{\epsilon},\bm\psi_{k})+\sum_{i=1}^{\beta}\nu(M,(i,i+2))\\
    = & {n \choose 2}+\beta,
\end{align*}
where we have used the fact that, by Fiedler's matrix tree theorem \cite{fiedler1975eigenvectors} (or inspection),  
$\sum_{k=1}^{n}\nu(M_{\epsilon},\bm\psi_{k})= {n \choose 2}$. 
 \end{proof}
 
Next, we build zero-diagonal matrices using a similar construction involving the path graph $P_{2n}$ on $2n$ vertices. The matrices constructed are nearly extremizers, with total nodal count equal to ${2n \choose 2 }+2\beta$ (rather than the lower bound ${2n \choose 2 }+\beta$ of Theorem \ref{thm: main bounds}).  

\begin{proposition}\label{prop: path 0diag}
Given $n \in \N$, $0 \le \beta \le 2(n-2)$, $\epsilon>0$ sufficiently small, and $\delta>0$ sufficiently small relative to $\epsilon$, let
$$M = D(\bm{u})  \otimes \begin{bsmallmatrix} 0 & 1 \\ 1& 0 \end{bsmallmatrix} - \epsilon A - \delta B \in \S(H^{(0)}_{2n,\beta}),$$
where $\bm{u} = -(n,n-1,\dots, 1)^T$, $D(\bm{u})$ is the diagonal matrix with $\bm{u}$ on the diagonal, $A$ is the adjacency matrix of the path $P_{2n}$ on $2n$ vertices, and $B$ is the adjacency matrix of $H^{(0)}_{2n,\beta}$. Let $\bm{\xi}_k \in \{\pm 1\}^n$, $k \in [n]$, be defined by
$$ \bm{\xi}_k(i) =\begin{cases} (-1)^{k-\ell} & \text{ for } \ell<k \\ \, +1 & 
			 \text{ for } \ell\ge k  \end{cases}.$$
The matrix $M$ satisfies Nodal Count Condition \ref{ass}, has eigenvectors $\bm{\varphi}_1,\ldots, \bm{\varphi}_{2n}$ with sign structure
$$ \mathrm{sign}(\bm{\varphi}_k) = \bm{\xi}_k \otimes \begin{bsmallmatrix} 1  \\ 1 \end{bsmallmatrix} \qquad  \text{and} \qquad \mathrm{sign}(\bm{\varphi}_{2n-k+1}) = \bm{\xi}_k \otimes \begin{bsmallmatrix} \phantom{-}1  \\ -1 \end{bsmallmatrix}$$ 
for all $k \in [n]$, and therefore has total nodal count
$$ {2n \choose 2} + 2\beta.$$
\end{proposition}

\begin{proof}
The proof is nearly identical to that of Proposition \ref{prop:path}: we first prove that $M_{\epsilon} := D(\bm{u})  \otimes \begin{bsmallmatrix} 0 & 1 \\ 1& 0 \end{bsmallmatrix} - \epsilon A$ has simple eigenvalues and non-vanishing eigenvectors with the desired $\bm \xi_{k}$ sign structure, and then show that $M=M_{\epsilon}-\delta B$ also has simple eigenvalues and non-vanishing eigenvectors with the same sign structure. The only difference in the two proofs lies in the proof of the non-vanishing eigenvectors and their sign structure for $M_{\epsilon}$, so we only prove this part. In fact, using that $M_{\epsilon}$ has zero diagonal and is supported on a bipartite graph, we later show that it is enough to prove that the first $n$ eigenvectors (among the $2n$) are non-vanishing eigenvectors with the desired sign structure. 

Consider $M_{\epsilon}=D(\bm{u})  \otimes \begin{bsmallmatrix} 0 & 1 \\ 1& 0 \end{bsmallmatrix} - \epsilon A$ as an $\epsilon$ perturbation of $D(\bm{u})  \otimes \begin{bsmallmatrix} 0 & 1 \\ 1& 0 \end{bsmallmatrix}$ for $\epsilon$ sufficiently small and fix $k\in[n]$. Let $\lambda_\epsilon = \sum_{i=0}^\infty \epsilon^i \lambda^{(i)}$ be the $k^{th}$ eigenvalue of $M_\epsilon$ and $\bm{\psi}_\epsilon = \sum_{i=0}^\infty \epsilon^i \bm{\psi}^{(i)}$ be the corresponding eigenvector, with \[\lambda^{(0)}=-(n-k+1) \quad \text{and}\quad \bm\psi^{(0)}= \bm{e}_k \otimes \begin{bsmallmatrix}  1 \\ 1 \end{bsmallmatrix},\]
which is the $k^{th}$ eigenpair of $D(\bm{u})  \otimes \begin{bsmallmatrix} 0 & 1 \\ 1& 0 \end{bsmallmatrix}$. For notational convenience, we use the indices 
\[x_{j}=2(k+j)-1 \quad \text{and} \quad y_{j}=2(k+j),\] 
so that $\bm \psi^{(0)}$ is supported on the interval $[x_{0},y_{0}]$, and $[2n]$ is the set of indices ranging from $x_{-(k-1)}$ to $y_{n-k}$. As in the proof of Proposition \ref{prop:path}, it suffices to prove the following claim for all $j$. \\ 

\noindent \textbf{Claim:} The vector $\bm\psi^{(j)}$ is supported inside the interval $[x_{-j},y_{j}]$, and
    \begin{align*}
      x_{-j} \in [2n]\; \Rightarrow & \quad \mathrm{sign}\big(\bm{\psi}^{(j)}(x_{-j}) \big) =\mathrm{sign}\big(\bm{\psi}^{(j)}(y_{-j}) \big) = (-1)^{j},\\
      y_{j} \in [2n]\; \Rightarrow & \quad  \mathrm{sign}\big(\bm{\psi}^{(j)}(x_{j}) \big) =\mathrm{sign}\big(\bm{\psi}^{(j)}(y_{j}) \big) =+1. \\
    \end{align*}

\noindent We proceed by induction on $j$. For $j=0$, $\bm\psi^{(0)}= \bm{e}_k \otimes \begin{bsmallmatrix}  1 \\ 1 \end{bsmallmatrix}$ and the claim holds. Now suppose that the desired claim holds for all vectors $\bm{\psi}^{(0)},\ldots,\bm{\psi}^{(j-1)}$. Equation \eqref{eq: pert vj full} gives the recursive formula
\[\bm{\psi}^{(j)}=\big(D(\bm{u})\otimes \begin{bsmallmatrix} 0 & 1 \\ 1& 0 \end{bsmallmatrix}-\lambda^{(0)}\big)^{+}\left[A \bm{\psi}^{(j-1)} + \sum_{i = 1}^j \lambda^{(i)} \bm{\psi}^{(j-i)}\right]\]
By our inductive hypothesis, the term $\sum_{i = 1}^j \lambda^{(i)} \bm{\psi}^{(j-i)}$ is supported inside the interval $[x_{-j+1},y_{j-1}]$, and $A \bm{\psi}^{(j-1)}$ is supported inside the larger interval $[x_{-j+1}-1,y_{j-1}+1]=[y_{-j},x_{j}]\subset [x_{-j},y_{j}]$. Since $\big(D(\bm{u})\otimes \begin{bsmallmatrix} 0 & 1 \\ 1& 0 \end{bsmallmatrix}-\lambda^{(0)}\big)^{+}$ is block diagonal with respect to the blocks $\{x_{i},y_i\}$, $i = -(k-1),\ldots,n-k$, $\bm\psi^{(j)}$ is supported inside $[x_{-j},y_{j}]$. Denote the $\{x_{i},y_{i}\}$ block of $\big(D(\bm{u})\otimes \begin{psmallmatrix} 0 & 1 \\ 1& 0 \end{psmallmatrix}-\lambda^{(0)}\big)^{+}$ by $C_{i}$ and let $r = -\lambda^{(0)}=n-k+1>0$. We have
\[C_{j}=\frac{1}{r^2- (r-j)^2} \begin{pmatrix} r & r- j \\ r- j& r \end{pmatrix} \quad \text{for } j \ne 0. \]
Now, if $x_{-j}\in [2n]$, then 
\begin{align*}
    \mathrm{sign} \left(\begin{bmatrix}
        \bm\psi^{(j)}(x_{-j})\\
        \bm\psi^{(j)}(y_{-j})
    \end{bmatrix}\right) &=  \mathrm{sign}\left(C_{-j}    \begin{bmatrix}
        [A\bm\psi^{(j-1)}](x_{-j})\\
        [A\bm\psi^{(j-1)}](y_{-j})
    \end{bmatrix}\right)\\
    &= \mathrm{sign}(\bm\psi^{(j-1)}(x_{-j+1}))\, \mathrm{sign}\left(C_{-j}    \begin{bmatrix}
        0\\
        1
    \end{bmatrix}\right)\\
    &= (-1)^{j-1} \mathrm{sign}\left(\frac{1}{r^2 - (r+j)^2} \begin{bmatrix}  r+ j \\  r \end{bmatrix}\right)=(-1)^{j} \begin{bmatrix}
        1\\
        1
    \end{bmatrix},
\end{align*}
where we used the inductive hypothesis that $\mathrm{sign}(\bm\psi^{(j-1)}(x_{-j+1}))=(-1)^{j-1}$. Now suppose $y_{j}\in[2n]$. Then $r>r-j>0$ and, by our inductive hypothesis, $\mathrm{sign}\big([A\bm\psi^{(j-1)}](x_{j}) \big)=1$ and $[A\bm\psi^{(j-1)}](y_{j})=0$, so  
\begin{align*}
    \mathrm{sign}\left(\begin{bmatrix}
        \bm\psi^{(j)}(x_{j})\\
        \bm\psi^{(j)}(y_{j})
    \end{bmatrix}\right) &= \mathrm{sign}\left(C_{j}    \begin{bmatrix}
        1\\
        0
    \end{bmatrix}\right)\\
    &= \mathrm{sign}\left(\frac{1}{r^2-(r-j)^2 } \begin{bmatrix}  r \\  r-j \end{bmatrix}\right)= \begin{bmatrix}
        1\\
        1
    \end{bmatrix}.
\end{align*}
This proves our inductive step and, hence, our desired claim holds for all $j$. Therefore, if $\bm\psi_{1},\ldots,\bm\psi_{2n}$ are the eigenvalues of $M_{\epsilon}$, then $\mathrm{sign}(\bm{\psi}_k) = \bm{\xi}_k \otimes \begin{psmallmatrix} 1  \\ 1 \end{psmallmatrix}$ for all $k\in[n]$. The sign structure for the other $n$ eigenvectors follows from noting that $M_\epsilon$ has zero diagonal and is strictly supported on a bipartite graph with bipartition $\{1,3,\ldots,2n-1\}$ and $\{2,4,\ldots,n\}$. This implies that $\bm\psi_{2n-k+1}$, the eigenvector for the $k^{th}$ largest eigenvalue, has sign structure $ \bm{\xi}_k \otimes \begin{psmallmatrix} 1  \\ -1 \end{psmallmatrix}$. Thus we have proved that $M_{\epsilon}$ has simple eigenvalues with the desired sign structure.

What remains is to consider the total nodal count. As in the proof of Proposition \ref{prop:path}, for $\delta$ sufficiently small, the eigenvectors $\bm\varphi_{1},\ldots,\bm\varphi_{2n}$ of $M$ have the same sign structure as $\bm\psi_{1},\ldots,\bm\psi_{2n}$. The total nodal count of $M$ equals the total nodal count of $M_{\epsilon}$, which is ${2n \choose 2}$, plus $\sum_{i=1}^{\beta}\nu(M,(i,i+4))$ (see Equation \eqref{eq:ENC} for definition). The calculated sign structure has the property that
\[\bm\varphi_{k}(i)\bm\varphi_{k}(i+4)<0 \quad \text{if and only if} \quad k = \lceil i/2 \rceil +1 \ \text{or} \ k = 2n - \lceil i/2 \rceil.\] 
Therefore, $\nu(M,(i,i+4))=2$ for every $(i,i+4) \in E(H^{(0)}_{2n,\beta})$, and the total nodal count equals
  \[{2 n \choose 2} + 2\beta.\]
	\end{proof}
	\begin{remark}\label{rem:NCC on P2n}
	In the proof of Proposition \ref{prop: path 0diag}, we provided a zero-diagonal matrix $M_{\epsilon} := D(\bm{u})  \otimes \begin{bsmallmatrix} 0 & 1 \\ 1& 0 \end{bsmallmatrix} - \epsilon A \in\mathcal{S}_{0}(P_{2n})$ that satisfies the Nodal Count Condition \ref{ass}. The existence of such a matrix is a key ingredient in the proof of Lemma \ref{lm:cycle}, a crucial lemma in the proof of Theorem \ref{thm: NCC characterization intro}.
	\end{remark}

	\subsection{Nodal count for bipartite graphs}
 Here we show, in contrast to Theorem \ref{thm: main bounds}, that if $G$ is bipartite and $A \in \mathcal{S}_0(G)$ satisfies (NCC), then its nodal count per edge is exactly $n/2$, implying that its average nodal count is exactly $\frac{|E(G)|}{2}=\frac{n-1}{2}+\frac{\beta(G)}{2}$.
	\begin{proposition}\label{prop:biPartite}
		Suppose $G = ([n],E)$ is bipartite, $n$ is even, and $A\in\S_{0}(G)$ satisfies Nodal Count Condition \ref{ass}. Then $\nu(A,(ij))=n/2$ for any edge $(i,j)\in G$ and the average nodal count is
		\[\frac{1}{n}\sum_{(ij)\in E(G)}\nu(A,(ij))=\frac{ |E(G)|}{2}=\frac{n-1}{2}+\frac{\beta(G)}{2}.\]
	\end{proposition}

	\begin{proof}
 
		Since $G$ is bipartite, there exists $u\in\{-1,1\}^{n}$ such that $u_{i}u_{j}=-1$ for any edge $(i,j)\in G$. If $A\in\S_{0}(G)$, then $A_{ij}u_{i}u_{j}=-A_{ij}$ for any $i,j$. Let $U=\diag (u)$ so that it is real orthogonal with $U^{-1}=U$, and $UAU=-A$. In particular, if $v$ is an eigenvector with eigenvalue $\lambda$, then $Uv$ is an eigenvector with eigenvalue $-\lambda$. Suppose that $A$ satisfies (NCC), then all the eigenvalues are simple and come in pairs of $\pm\lambda$, and since there is an even number of them then $0$ is not an eigenvalue of $A$. So, for any eigenvector $v$, the vector $Uv$ is a different eigenvector and $\mathrm{sign}(Uv(i)Uv(j))=-\mathrm{sign}(v(i)v(j))$, $\nu(i,j,v)+\nu(i,j,Uv)=1$ for every edge $(i,j)\in G$. It follows that $\nu(A,(ij))=n/2$ for every edge.   
	\end{proof}

  \begin{remark}
     Note that a bipartite $G$ with an $A\in\S_{0}(G)$ that satisfies (NCC) must have a perfect matching, and hence $n$ is even, otherwise $\det(A)=0$ for all $A\in\S_{0}(G)$ (see \cite[p. 307]{lovasz2009matching} for example), in which case Theorem \ref{thm: NCC charecterization} says any $A\in\S_{0}(G)$ fails to satisfy (NCC).
 \end{remark}
 
		\section{The graph characterization of the Nodal Count Condition for $\S_{0}(G)$}\label{sec:ncc}
	
	Constant diagonal matrices are quite important in practice, as this class includes the adjacency matrix and normalized Laplacian matrix of a graph and the unnormalized Laplacian of a regular graph. However, unlike for $\S(G)$, where it is known that for any $G$ graph a generic matrix in $\S(G)$ satisfies (NCC), the case of matrices in 
$\S_{0}(G)$ satisfying (NCC) is more subtle. For instance, there exist graphs for which no matrix in $\S_{0}(G)$ satisfies (NCC). The following theorem is a restatement of Theorem \ref{thm: NCC characterization intro} and provides a full characterization. Recall Definition \ref{def:subdeterminantal}: a graph $G$ is said to be \emph{sub-determinantal} if $\det(A)=0$ for all $A\in\mathcal{S}_{0}(G)$, and said to be \emph{determinantal} otherwise.  
\begin{theorem}\label{thm: NCC charecterization}
    For any simple connected graph $G$ exactly one of the following two holds:
    \begin{enumerate}
        \item If $G$ is determinantal, then the set of matrices in $\S_{0}(G)$ that fail to satisfy Nodal Count Condition \ref{ass} is a closed (semi-algebraic) set of positive co-dimension. 
        \item If $G$ is sub-determinantal, then any $A\in\S_{0}(G)$ fails to satisfy Nodal Count Condition \ref{ass} in one of the following ways:
          \begin{enumerate}
      \item Either $\lambda=0$ is a multiple eigenvalue of $A$, or
      \item $\lambda = 0$ is a simple eigenvalue and the corresponding eigenvector $\bm\varphi$ satisfies $\bm\varphi(i)\bm\varphi(j)=0$ for every edge $(ij) \in E(G)$. 
  \end{enumerate}
    \end{enumerate}
\end{theorem}

We prove this theorem by decomposing it into the following four lemmas. 
	\begin{lemma}\label{lem: genericity}
	    For any simple connected graph $G$, either all matrices in $\S_{0}(G)$ fail to satisfy Nodal Count Condition \ref{ass}, or it is satisfied generically. That is, if $B\subset\S_{0}(G)$ is the set of matrices that fail to satisfy Nodal Count Condition \ref{ass}, then either $B=\S_{0}(G)$ or $B$ is a closed semi-algebraic set of positive co-dimension.   
	\end{lemma}
 \begin{proof}
Assuming that there exists some  $A'\in\mathcal{S}_{0}(G) \backslash B$, it is enough to show that the set $B$ is a closed semi-algebraic set of positive co-dimension. 

 Recall that the entries of the adjugate matrix $\mathrm{adj}(A-\lambda I)$ are (up to signing) the minors of $A-\lambda I$, and are therefore polynomials in $A$ and $\lambda$. The rank of $\mathrm{adj}(A-\lambda I)$ is $n$ (full-rank) when $\lambda$ is not an eigenvalue, it is the zero matrix $\mathrm{adj}(A-\lambda I)=0$ when $\lambda$ is a repeated eigenvalue, and it has rank one if $\lambda$ is a simple eigenvalue. In particular, if $\lambda$ is simple with normalized eigenvector $\bm\varphi$, then $\mathrm{adj}(A-\lambda I)=c\bm\varphi \bm\varphi ^{T}$ for a non-zero scalar $c\ne 0 $. This means that 
 \[B=\{A\in\mathcal{S}_{0}(G) \, | \, \exists \, k,i\in[n]\ \text{s.t.}\  \mathrm{adj}(A-\lambda_{k}(A) I)_{ii}=0\}.\]
It is clear that $B$ is closed, since $A\mapsto\lambda_{k}(A)$ is continuous, so $A\mapsto\mathrm{adj}(A-\lambda_{k}(A) I)_{ii}$ is continuous for any $i,k\in[n]$, and $B$ is the union of their zero sets. By Tarski-Seidenberg, $B$ is also semi-algebraic. To be more specific, consider the algebraic sets
     \begin{align*}
     X= & \{(A,\lambda)\in\S_{0}(G)\times\R \, | \, \det(A-\lambda I)=0 \},\\
     X_{\mathrm{mult}}= & \{(A,\lambda)\in X\, | \, \mathrm{discriminant}(A)=0 \},\\ X_{i}= & \{(A,\lambda)\in X\, | \, \mathrm{adj}(A-\lambda I)_{ii}=0 \}\ \text{for all }\  i\in[n].    
     \end{align*}
     Consider the projection $\pi:X\to\S_{0}(G)$, so that \[B_{\mathrm{mult}}:=\{A\in \S_{0}(G)\, | \, \mathrm{discriminant}(A)=0 \}=\pi(X_{\mathrm{mult}})\] is algebraic, and $B_{i}=\pi(X_{i})$ is semi-algebraic by Tarski-Seidenberg, and therefore so is $B=B_{\mathrm{mult}} \cup [\cup_{i=1}^{n}B_{i}]$. By assumption, there exists $A' \in \mathcal{S}_0(G) \backslash B$, so $B_{\mathrm{mult}}$ has positive codimension as an algebraic strict subset. What remains is to show that $B_{i}$ has positive codimension. Assume, for the sake of contradiction, that this is not the case for some $B_{i}$. Then, there is an open neighborhood $O\subset B_{i}\setminus B_{\mathrm{mult}} $, and up to going to a sub-neighborhood, there is $k\in[n]$ such that $\mathrm{adj}(A-\lambda_{k}(A) I)_{ii}=0$ for all $A\in O$. Let $A\in O$ and $A'\notin B$ as assumed, and consider the one-parameter family $A_{t}=tA'+(1-t)A$ for $t\in[0,1]$. Then there is an analytic function $\lambda(t)$, which is an eigenvalue of $A_{t}$ for all $t$ and equals $\lambda_{k}(A_{t})$ for small $t$, by Lemma \ref{lm: katto-rellich}, and so $t\mapsto\mathrm{adj}(A-\lambda(t) I)_{ii}=0$ is the zero function, and in particular $\mathrm{adj}(A'-\lambda(1) I)_{ii}=0$ and $\det(A'-\lambda(1) I)=0$, in contradiction to our assumption.    
 \end{proof}

\begin{lemma}\label{lm:cycle}
If $G=C_{2n+1}$ is an odd cycle, then a generic matrix in $\S_0(G)$ satisfies Nodal Count Condition \ref{ass}.
\end{lemma}

\begin{proof}
By Lemma \ref{lem: genericity} it is enough to construct a matrix $A\in\S_0(G) $ that satisfies (NCC). We do so by making use of a matrix $\hat A \in \S_0(P_{2n})$, where $P_{2n}$ is the path on $2n$ vertices, with $\det(\hat A) \ne 0$ that satisfies (NCC); we already constructed such a matrix in Subsection \ref{sub:treeperturb} (see Remark \ref{rem:NCC on P2n}). Consider the matrix
\[ A_\epsilon = A + \epsilon B \in \S_0(C_{2n+1}), \qquad A = \begin{blockarray}{ccc}
{\scriptstyle 2n} & {\scriptstyle 1} & \\
\begin{block}{[cc]c}
\hat A & 0 & {\scriptstyle 2n} \\  0 & 0 & {\scriptstyle 1} \\
    \end{block}
\end{blockarray}, \qquad  B = \begin{blockarray}{ccc}
{\scriptstyle 2n} & {\scriptstyle 1} & \\
 \begin{block}{[cc]c}
 0 & \bm{e}_1 + c  \bm{e}_{2n} & {\scriptstyle 2n} \\   \bm{e}^T_1 + c  \bm{e}^T_{2n} & 0 & {\scriptstyle 1} \\
 \end{block}
\end{blockarray},\]
where $c$ is a non-zero constant that is not equal to $-\bm \varphi (1)/\bm \varphi(2n)$ for any eigenvector $\bm \varphi$ of $\hat A$. The matrix $A_\epsilon$ is strictly supported on $\S_0(C_{2n+1})$. Lemma \ref{lm: perturbation general} provides a small interval $ (0,b) $ so that, for all $ \epsilon\in(0,b) $, the eigenvalues of $A_{\epsilon}$ are simple and the support and signs of the eigenvectors are independent of $ \epsilon $. In order to prove $A_\epsilon$, $\epsilon \in (0,b)$, satisfies (NCC), all that remains is to show that its eigenvectors do not vanish at an entry. For $ \epsilon\in(0,b) $, an eigenpair of $A_{\epsilon}$ has the form  $\lambda_{\epsilon}=\lambda^{(0)}+\epsilon \lambda^{(1)} + O(\epsilon^2) $ and $ \bm\varphi_{\epsilon}=\bm\varphi^{(0)}+\epsilon \bm\varphi^{(1)}+O(\epsilon^{2})$. We have $\lambda^{(1)} = \langle B \bm\varphi^{(0)},\bm{\varphi}^{(0)} \rangle = 0$ for all eigenvectors, simplifying the analysis. We treat two cases, depending on whether $\lambda^{(0)} =0$.

First, consider an arbitrary eigenpair of $\hat{A}$, say $(\lambda_{\epsilon}, \bm\varphi_{\epsilon})$, where $\lambda^{(0)} \ne 0$. As $\mathrm{sign}(\bm\varphi_{\epsilon})$ is independent of $\epsilon$ in this region, and $\bm\varphi^{(0)}(i) \ne 0$ for all $i \ne 2n+1$, it suffices to consider only $\bm\varphi^{(1)}(2n+1)$. By Lemma \ref{lm: perturbation simple}, $$\bm\varphi^{(1)} = - (A -\lambda^{(0)}I)^+ B \bm\varphi^{(0)},$$
and so
$$\bm\varphi^{(1)}(2n+1) = \frac{1}{\lambda^{(0)}} (\bm\varphi^{(0)}(1) + c \bm \varphi^{(0)}(2n)) \ne 0.$$

Next, consider the case $(\lambda_{\epsilon}, \bm\varphi_{\epsilon})$, where $\lambda^{(0)} = 0$. In this setting $\bm\varphi^{(0)} = \bm{e}_{2n+1}$, and, again, by Lemma \ref{lm: perturbation simple}, $\bm\varphi^{(1)} = - A^+ B \bm e_{2n+1}$. This implies that
$$ \bm\varphi^{(1)}(i)= \hat A^{-1}_{i,1}+c\hat A^{-1}_{i,2n}\quad \text{ for all } i\in[2n].$$
By Proposition \ref{prop:tree_inverse}, $\hat A^{-1}_{i,1} \ne 0$ if and only if $i$ is even, and $A^{-1}_{i,2n} \ne 0$ if and only if $i$ is odd, so $ \bm\varphi^{(1)}(i)\ne0$ for all $i\in[2n]$, completing the proof.
\end{proof}
 
 	\begin{lemma}
	    If $G$ is determinantal, namely, there exists $A'\in\mathcal{S}_{0}(G)$ with $\det(A')\ne0$, then there is a (vertex) disjoint union of edges and odd cycles of $G$ that cover all vertices, and a generic matrix in $\S_0(G)$ satisfies Nodal Count Condition \ref{ass}. 
	\end{lemma}

\begin{proof}
By Lemma \ref{lem: genericity} it is enough to construct a matrix $A_{\epsilon}\in\S_0(G) $ that satisfies (NCC), which we do by a clever choice of $A$ with simple eigenvalues  and $B$ such that we can show $A_{\epsilon}=A+\epsilon B$ satisfies (NCC). In fact, it is enough to construct a matrix $A_{\epsilon}\in\S_0(G) $ with simple eigenvalues and non-vanishing eigenvectors that is only strictly supported on a subgraph of $G$, as we can always perturb again by taking $A_{\epsilon,\delta}=A_{\epsilon}+\delta B'$ with $B'$ being the adjacency matrix of $G$, so that for small enough $\delta>0$,  $A_{\epsilon,\delta}\in\S_0(G)$ is strictly supported on $G$, and satisfies (NCC) by Lemma \ref{lm: perturbation general} and Lemma \ref{lm: perturbation simple}. 
 
 Now, because there exists an $A'\in\S_0(G) $ with $\det(A') \ne 0$, at least one term in the Leibniz expansion of the determinant is non-zero, i.e., $\prod_{j=1}^{n}A'_{j,\sigma(j)} \ne 0$ for some permutation $\sigma $. Because $A'$ has zero-diagonal, $\sigma$ has no $1$-cycles (i.e., $\sigma$ is a dearrangement). By associating each cycle of $\sigma$ of length $k$ with the analogous undirected cycle subgraph of $G$ of length $k$ (where a $2$-cycle is simply an edge), $\sigma$ encodes a vertex-disjoint cycle cover of $G$ (where an edge is considered a $2$-cycle). Any disjoint cycle cover can be converted to one with only odd cycles and edges, as every cycle of even length can be disjointly covered using alternating edges. Therefore, $G$ has a vertex-disjoint cycle cover consisting only of edges and odd cycles. Let us consider one such cover $H_1,\ldots,H_s$, and denote by $V_1,\ldots,V_s$ the corresponding partition of the vertex set $[n]$ of $G$, where each subset $V_i$ corresponds to the vertex set of $H_i$ (where, again, $H_i$ either an edge or an odd cycle).
 
 We are now prepared to construct a matrix $A_\epsilon \in \S_0(G)$ that satisfies (NCC) using the cycle cover associated with $V_1,\ldots,V_s$. Let $A$ be block diagonal with respect to $V_1,\ldots,V_s$, and denote the matrix $A$ restricted to the entries of $V_i$ by $A_i$. We require that each $A_i \in \S_0(H_i)$, $i = 1,\ldots, s$, satisfies (NCC) and has $\det(A_i) \ne 0$, and that the matrices have disjoint spectra, i.e., $\Lambda(A_i) \cap \Lambda(A_j) = \emptyset$ for all $i \ne j$. For a single edge, we note that $\begin{psmallmatrix}
     0 & 1 \\ 1& 0 
 \end{psmallmatrix}$ satisfies (NCC) and has non-zero determinant, and Lemma \ref{lm:cycle} provides such a matrix for an odd cycle of any length. Disjoint spectra can be obtained simply by properly scaling each $A_i$, as none of these matrices has a zero eigenvalue. Furthermore, by properly scaling each $A_i$, we may assume that each pair $\lambda$ and $A_i$, $\lambda \in \Lambda(A_j)$, $j \ne i$, is such that $\lambda$ is not in the finite set of Proposition \ref{prop:inverse_sparsity} associated with $A_i$.

 Let $\tilde E \subset E(G)$ denote a minimal set of edges needed to be added to the cover $H_1+H_2 +\ldots +H_s$ so that
 \[\tilde H = ([n],\tilde E \cup E(H_1) \cup \dots \cup E(H_s))\] is a connected subgraph of $G$. One important consequence of the minimality of $\tilde E$ is that, for any path $P_{ij}:= v_0 \, v_1 \, \dots \, v_k$ between $H_i$ and $H_j$ (i.e., $v_0 \in V_i$, $v_k \in V_j$, and $v_p \not \in V_i \cup V_j$ for $ 1<p<k$), $v_0$, $v_k$, and $E(P_{ij}) \cap \tilde E$ are independent of the choice of path. Therefore, we may define the distance $d_H(V_i,V_j)$ between $H_i$ and $H_j$ to be $|E(P_{ij}) \cap \tilde E|$ for a path $P_{ij}$ between $H_i$ and $H_j$ in $\tilde H$, and we may speak of a unique last vertex $v_k \in V_j$ of $P_{ij}$. Let $B$ be the binary matrix with $B_{ij} =1$ if and only if $(ij) \in \tilde E$. We define $A_\epsilon = A + \epsilon B$.

By construction, $A$ has simple eigenvalues, and so, by Lemma \ref{lm: perturbation general}, there is a small interval $ (0,b) $ so that for all $ \epsilon\in(0,b) $, the eigenvalues of $A_{\epsilon}$ are simple and the support and signs of the eigenvectors are independent of $ \epsilon $. What remains is to prove that the eigenvectors of $A_{\epsilon}$ do not vanish at a vertex. Consider an arbitrary eigenpair of $A$ and extend it to an eigenpair of $A_{\epsilon}$ with a power series $\lambda_\epsilon = \sum_{i=0}^\infty \epsilon^i \lambda^{(i)}$ and $\bm{\varphi}_\epsilon = \sum_{i=0}^\infty \epsilon^i \bm{\varphi}^{(i)}$. Because $A$ is block diagonal with simple eigenvalues, $\bm{\varphi}^{(0)}$ is strictly supported on a single block, without loss of generality, given by $V_1$. According to Lemma \ref{lm: perturbation general}, it is enough to show that for every vertex $i$ there is some $\ell$ such that $\bm{\varphi}^{(\ell)}(i)\ne0$. We will do so by inducting on the subsets $V_1,\ldots,V_s$ according to distance $d_H(V_1,V_j)$. Let $J_0 = V_1$, $J_{r} \subset [n]$ equal the union of all $V_j$ at distance $d_H(V_1,V_j) = r$, and $J_{\le R }=\cup_{r=0}^{R}J_{r}$.

To complete the proof, we prove by induction that $\bm{\varphi}^{(\ell)}$ is non-zero on the vertex set $J_\ell$ and supported inside the ball $J_{\le \ell}$. By construction, this is true for $\ell = 0$. For our inductive step, suppose that our statement holds for all $\ell'< \ell$. Because $\langle B\bm\varphi^{(0)},\bm\varphi^{(0)}\rangle=0$, Equation \eqref{eq: pert vj} provides a recursive formula for $\bm\varphi^{(\ell)}$ (when $\ell=1$ neglect the sum)
\begin{equation*}
\bm\varphi^{(\ell)}=-(A-\lambda^{(0)} I)^{+}B\bm\varphi^{(\ell-1)}+\sum_{m=2}^{\ell}\lambda^{(m)}(A-\lambda^{(0)} I)^{+}\bm\varphi^{(\ell-m)}.  
\end{equation*}
We note that $(A-\lambda^{(0)} I)^{+}$ is block-diagonal with respect to $V_1,\ldots,V_s$, and so $(A-\lambda^{(0)} I)^{+}\bm{\varphi}^{(\ell')}$ is supported in the ball $J_{\le \ell'}$ for all $\ell'<\ell$. Therefore, 
\[\bm\varphi^{(\ell)}|_{J_{r}}=-\left((A-\lambda^{(0)} I)^{+}B\bm\varphi^{(\ell-1)}\right)|_{J_{r}} \qquad \text{ for all }\ r\ge \ell. \]
Because the matrix $B$ is non-zero precisely on the edge set $\tilde E$, $B\bm\varphi^{(\ell-1)}$ is supported on $J_{\le \ell}$ and so $\bm\varphi^{(\ell)}|_{J_{r}}$ is zero for $r > \ell$. What remains is to prove that $\bm\varphi^{(\ell)}|_{J_{\ell}}$ does not vanish at a vertex.

Consider an arbitrary $V_j$ in $J_\ell$, and let $\tilde v \in V_j$ denote the unique last vertex of every path $P_{1j}$ from $V_1$ to $V_j$ with interior vertices not in $V_1 \cup V_j$ (as described above). Then $B\bm\varphi^{(\ell-1)}|_{V_{j}}$ is non-zero at $\tilde v$ and equal to zero elsewhere. By Proposition \ref{prop:inverse_sparsity} and our choice of scaling for each $A_i$, $(A_j-\lambda^{(0)} I)^{-1}$ is entry-wise non-zero. Therefore, $\bm\varphi^{(\ell)}|_{J_{r}}$ is non-vanishing, completing the proof.
 \end{proof}

 \begin{lemma}
     If $G$ is sub-determinantal, namely, $\det(A)=0$ for all $A\in\S_{0}(G)$, then, for any $A\in\S_{0}(G)$,
     \begin{enumerate}
         \item Either $\lambda=0$ is a multiple eigenvalue, or
         \item $\lambda = 0$ is a simple eigenvalue and the corresponding eigenvector $\bm\varphi$ satisfies $\bm\varphi(i)\bm\varphi(j)=0$ for every edge $(ij) \in E(G)$.
     \end{enumerate}
 \end{lemma}
 \begin{proof}
     Suppose $\det(A)=0$ for all $A\in\S_{0}(G)$. Let 
$(ij) \in E(G)$ and $E_{ij}$ be the symmetric matrix with entries $E_{ij}=E_{ji}=1$ and zero elsewhere. Jacobi's formula for the derivative of the determinant gives \[\frac{\partial}{\partial t}\det(A+tE_{ij})|_{t=0}=\mathrm{trace}(E_{ij}\mathrm{adj}(A))=(\mathrm{adj}(A))_{ij}+(\mathrm{adj}(A))_{ji}=2(\mathrm{adj}(A))_{ij}.\] Since $A+tE_{ij}\in\S_{0}(G)$ for all $t$, $t\mapsto\det(A+tE_{ij})\equiv0$ and so $\frac{\partial}{\partial t}\det(A+tE_{ij})|_{t=0}=2(\mathrm{adj}(A))_{ij}=0$. Since $\det(A)=0$, this means that either $\lambda=0$ is multiple eigenvalue, or it is simple with an eigenvector that satisfies $(\bm\varphi\bm\varphi ^{T})_{ij}=0$. 
 \end{proof}
 \begin{corollary}
     Let $G$ be a bipartite graph (any tree for example). If $G$ has no perfect matching, then any $A\in\S_{0}(G)$ fails to satisfy Nodal Count Condition \ref{ass}.   
 \end{corollary}
 \begin{proof}
 This follows from the fact that the determinant of any $A\in\S_{0}(G)$ is zero if $G$ has no perfect matching. See \cite[p. 307]{lovasz2009matching} for example.
 \end{proof}

	\section{Vanishing Eigenvectors and Eigenvalue Multiplicity}\label{sec:examples}
	
	The set of matrices in $\S(G)$ that fails to satisfy (NCC) is an algebraic subvariety of positive co-dimension, as stated in Theorem \ref{thm: NCC characterization intro}, implying that any matrix $A \in \mathcal{S}(G)$ with either a repeated eigenvalue or an eigenvector that vanishes at a vertex is arbitrarily close to another matrix in $\S(G)$ that does satisfy (NCC). However, due to the fact that many matrices in practice fail to satisfy this condition, a great deal of research has been devoted to understanding what happens in this case, particularly regarding nodal domains (see, for instance, \cite{BRIANDAVIES200151,davies2000discrete,gladwell2002courant,mckenzie2023nodal,urschel2014spectral,urschel2018nodal,xu2012nodal} and the references cited therein). Here we make a number of brief observations regarding the total nodal count when (NCC) fails. We break our analysis into three cases: matrices with an eigenvector that vanishes at a vertex, matrices with a repeated eigenvalue, and, finally, matrices with ``both'', by which we mean that there is a repeated eigenvalue with no nowhere-vanishing eigenvector.
 \begin{remark}
     If a matrix has a repeated eigenvalue $\lambda$ with eigenspace $V$, for any $i\in[n]$ there is always a
$\bm\varphi\in V$ with $\bm\varphi(i)=0$. There is a nowhere-vanishing $\bm\varphi\in V$ if and only if for any $i\in[n]$ there is some $\bm\varphi'\in V$ with $\bm\varphi'(i)\ne 0$. Therefore, in the presence of multiplicity, the relevant question is not whether a matrix admits some ``bad" eigenbasis, which is always the case, but, rather, whether there exists a ``good" (i.e., nowhere-vanishing) eigenbasis.
 \end{remark}
	
	\subsection{Vanishing eigenvectors} Suppose $A \in \mathcal{S}(G)$ has simple eigenvalues $\lambda_1<\ldots<\lambda_n$, but there exists an eigenvector $\bm{\varphi}_k$ with $\bm{\varphi}_k(i) = 0$ for some $i \in [n]$. In this case, what constitutes an edge where an eigenvector changes sign is ambiguous, and is not necessarily covered by Theorem \ref{thm: main bounds}. We give the following illustrative example.
	
	\begin{example}[bounds fail for weak/strong formulation]\label{ex: vanish}
		The matrix 
  \[A = \begin{bmatrix} \phantom{-}0 & -1 & \phantom{-}0 & \phantom{-}0 \\  -1 & \phantom{-}0 & -1 & -1 \\ \phantom{-}0 & -1 & \phantom{-}1 &-1 \\ \phantom{-}0 & -1 & -1 & \phantom{-}1 \end{bmatrix}\]
  has eigenvalues $- \sqrt{3}$, $0$, $\sqrt{3}$, $2$, with corresponding eigenvectors  
  \[ \begin{bmatrix} 1 \\ \sqrt{3} \\ 1 \\ 1 \end{bmatrix}, \; \begin{bmatrix} \phantom{-}2 \\ \phantom{-}0 \\ -1 \\ -1 \end{bmatrix}, \; \begin{bmatrix} \phantom{-}1 \\ -\sqrt{3} \\ \phantom{-}1 \\ \phantom{-}1 \end{bmatrix}, \; \begin{bmatrix} \phantom{-}0 \\ \phantom{-}0 \\ -1 \\ \phantom{-}1 \end{bmatrix}.\]
		If $A$ had non-vanishing eigenvectors, Theorem \ref{thm: main bounds} would imply that the total nodal count would be between $7$ and $9$. However, if we count the total number of edges where $A_{ij} \bm{\varphi}_k(i) \bm{\varphi}_k(j) > 0$ over all $k$, there are only $4$. Counting edges where $A_{ij} \bm{\varphi}_k(i) \bm{\varphi}_k(j) \ge 0$ also does not fix this issue, as we obtain a total count of $10$.
	\end{example}
	
	This issue is due to the fact that a vertex with a value of zero may simultaneously count as a ``positive" vertex with respect to some neighbors and a ``negative" vertex to others, leading to abnormally large nodal counts. However, because this property is of positive co-dimension, we can immediately make the following claim (in the spirit of \cite[Theorem 3.6]{urschel2018nodal}, \cite[Theorem 1.3]{mckenzie2023nodal}).
	
	\begin{proposition}\label{prop: signing}
    Suppose $A\in\mathcal{S}(G)$ is strictly supported on $G$ and has simple eigenvalues with corresponding eigenvectors $ \bm{\varphi}_1,\ldots,\bm{\varphi}_n$. Then there exists a choice of signings $ \bm{\xi}_1,\ldots,\bm{\xi}_n \in \{\pm 1 \}^{n}$
    with $\mathrm{sign}(\bm{\xi}_k(i))=\mathrm{sign}(\bm{\varphi}_k(i))$ for all $\bm{\varphi}_k(i)\ne0$ for which the total nodal count $\sum_{k=1}^{n}\nu(A,\bm{\xi}_k)$ satisfies the bounds of Theorem \ref{thm: main bounds}.
	\end{proposition}
 \begin{proof}
     Given a matrix $B\in \mathcal{S}(G)$, the eigenvalues and eigenvectors of $A_{\epsilon
     }=A+\epsilon B$ are real analytic functions of $\epsilon\in\R$, so $A_{\epsilon}$ either fails to satisfy (NCC) for all $\epsilon\in \R$, or it satisfies (NCC) for all $\epsilon\in(0,T)$ for some $T>0$ depending on $B$. There must be some $B\in \mathcal{S}(G)$ for which the latter case holds, as the non-existence of such a matrix would contradict Theorem \ref{thm: NCC characterization intro}. Hence, we may choose some $B\in \mathcal{S}(G)$ such that $A_{\epsilon}$ satisfies (NCC) for all $\epsilon\in (0,T)$, for some $T>0$. Lemma \ref{lm: perturbation general} provides a possibly smaller interval $(0,b)$ such that the sign structure of the eigenvectors of $A_{\epsilon}$ is independent of $\epsilon\in (0,b)$ and agrees with that of the non-zero entries of $ \bm{\varphi}_1,\ldots,\bm{\varphi}_n$. We let  $\bm{\xi}_1,\ldots,\bm{\xi}_n$ equal the sign structure of the eigenvectors of $A_\epsilon$, $0<\epsilon< \min \{b,T\}$, implying that the total nodal count satisfies the bounds of Theorem \ref{thm: main bounds}, as $A_{\epsilon
     }$ satisfies (NCC).
 \end{proof}

	\subsection{Eigenvalue Multiplicity} When a matrix $A \in \mathcal{S}(G)$ has a repeated eigenvalue, e.g., $\lambda_k = \lambda_{k+1}$ for some $k$, there is no longer a notion of a total nodal count, as there is no longer a fixed eigenbasis. The question is then whether $A$ has a nowhere-vanishing eigenbasis, and if so, does its nodal count obey the bounds of Theorem \ref{thm: main bounds}. In this subsection, we provide two sufficient conditions on $A$ under which every choice of a non-vanishing eigenbasis satisfies the bounds of Theorem \ref{thm: main bounds}. However, we first describe a naive perturbative approach. If we perturb $A$ to a nearby $A'$ that satisfies (NCC), in which case the bounds of Theorem \ref{thm: main bounds} hold for $A'$, and we manage to do so without changing the sign structure of the chosen eigenbasis of $A$, then the initial eigenbasis must satisfy the bounds of Theorem \ref{thm: main bounds}. The caveat is, as seen in Lemma \ref{lm: perturbation general} part (1), that such a perturbation dictates the initial eigenbasis, not the other way around. To illustrate this problem, we show that diagonal perturbations, which are commonly used to resolve multiplicity, can produce sign structures that do not match that of any eigenbasis of the original matrix. For the resolution of eigenvalue multiplicity via diagonal perturbations see Minami-type estimates in the theory of Anderson localization \cite{combes2009generalized} or the DAD theorem of Garza-Vargas \cite{garza2023} for details. 

\begin{example}[diagonal perturbations can ruin eigenspace sign structure]\label{ex: cycle}
Let $A$ be the adjacency matrix of the cycle $C_{4}$ on $4$ vertices. The nullspace of $A$ 
\[\mathrm{null}(A) = \{ (a,b,-a,-b)^T \, | \, a,b \in \mathbb{R} \}\]
is of dimension two, and the vectors $\bm{x} = (1, 0 -1,0)^T/\sqrt{2}$ and $\bm{y} =  (0, 1,0,-1)^T / \sqrt{2}$ form an orthonormal basis for $\mathrm{null}(A)$. Consider the matrix $A_\epsilon = A + \epsilon D \in \mathcal{S}(C_{4})$, where $D = \mathrm{diag}(D_1,D_2,D_3,D_4)$, $\|D\|_2 = 1$, is a diagonal matrix, and $D_1 + D_3 \ne D_2 + D_4$. 
The restriction of $D$ to $\mathrm{null}(A)$ is diagonalized by $\bm{x}$ and $\bm{y}$, with
\[  \begin{bmatrix} \bm{x}^T \\ \bm{y}^T \end{bmatrix}    D \begin{bmatrix} \bm{x} & \bm{y} \end{bmatrix} = \begin{bmatrix} (D_1 + D_3)/2 & 0 \\ 0 &  (D_2 + D_4)/2  \end{bmatrix} ,\]
so, by Lemma \ref{lm: perturbation general} part (1) and Lemma \ref{lm: perturbation simple} part (3), 
$A_\epsilon$ has simple eigenvalues for all $\epsilon\in(0,b)$ for some $b>0$, with real analytic eigenvectors,  
\begin{align*}
\bm{x}_\epsilon &= \bm{x} + \epsilon \bigg[ A^+ \big( \tfrac{D_1 + D_3}{2} I - D \big) \bm{x} - \frac{\bm{x}^T D A^+ D \bm{y}}{\frac{D_1+D_3}{2} - \frac{D_2+D_4}{2}} \, \bm{y}\bigg] + O(\epsilon^2)  ,\\
\bm{y}_\epsilon &= \bm{y} + \epsilon \bigg[ A^+ \big( \tfrac{D_2 + D_4}{2} I - D \big) \bm{y} - \frac{\bm{x}^T D A^+ D \bm{y}}{\frac{D_2+D_4}{2} - \frac{D_1+D_3}{2}} \, \bm{x}\bigg] + O(\epsilon^2),
\end{align*}
of corresponding eigenvalues 
${}_{\bm{x}}\lambda_\epsilon = \epsilon (D_1+D_3)/2 + O(\epsilon^2)$, and ${}_{\bm{y}}\lambda_\epsilon = \epsilon (D_2+D_4)/2 + O(\epsilon^2)$. 
We have $A^+ = \frac{1}{4} A$, and, after a series of calculations, note that
\[\bm{x}_\epsilon = \frac{1}{\sqrt{2}}\begin{bmatrix} 1 \\ \frac{\epsilon(D_1-D_3)(2D_4 - D_1 - D_3)}{4(D_1-D_2+D_3-D_4)} \\ -1 \\ \frac{\epsilon(D_1-D_3)(2D_2 - D_1 - D_3)}{4(D_1-D_2+D_3-D_4)}  \end{bmatrix}  + O(\epsilon^2) \quad \text{and} \quad \bm{y}_\epsilon = \frac{1}{\sqrt{2}}\begin{bmatrix} \frac{\epsilon(D_2-D_4)(2D_3 - D_2 - D_4)}{4(-D_1+D_2-D_3+D_4)}\\ 1 \\ \frac{\epsilon(D_2-D_4)(2D_1 - D_2 - D_4)}{4(-D_1+D_2-D_3+D_4)}\\ -1   \end{bmatrix}  + O(\epsilon^2) .\]
For $D$ fixed and $\epsilon$ sufficiently small, the sign structure of $\bm{x}_\epsilon$ and $\bm{y}_\epsilon$ matches that of eigenvectors in $\mathrm{null}(A)$, i.e., $\mathrm{sign}(\bm{x}_\epsilon) = \mathrm{sign}(\bm\psi_1)$ and $\mathrm{sign}(\bm{y}_\epsilon) = \mathrm{sign}(\bm\psi_2)$ for some $\bm\psi_1,\bm\psi_2 \in \mathrm{null}(A)$, only if
\begin{align*}
    \mathrm{sign}(2D_4 - D_1 - D_3) &= -\mathrm{sign}(2D_2 - D_1 - D_3) \quad \text{and} \\
    \mathrm{sign}(2D_3 - D_2 - D_4) &= -\mathrm{sign}(2D_1 - D_2 - D_4).
\end{align*}
Without loss of generality, suppose that $D_1 = 1$ and $D_2>D_4$. By inspection, 
\[\mathrm{vol}(\{ (D_2,D_3,D_4) \in [-1,1]^3 \, | \, 2 D_4 \le 1 + D_3 \le 2 D_2 \text{ and } 2 D_3 \le D_2 + D_4 \le 2 \}) = 10/9,\]
and so the fraction of diagonal matrices $D$ with $\|D\|_2 =1$ that satisfies both of the above conditions is $5/18$, implying that $72.\bar{2}\%$ of the possible diagonal perturbations (measured w.r.t. the uniform measure on the cube $\|D\|_2 =1$) of $A$ do not preserve the sign structure of the nullspace.

\end{example}
 
\begin{remark}
    The above example can be generalized to any cycle on $4n$ vertices, with $n\in\N$, in which case 
    \[\mathrm{null}(A) = \{ \bm{\varphi} \in \mathbb{R}^{4n} \, | \, \bm{\varphi}(i) = -\bm{\varphi}(j) \text{ for all } i \equiv 0, j \equiv 2 \mod 4 \text{ and } i \equiv 1, j \equiv 3  \mod 4 \},\]
    and, for any diagonal $D$, the restriction $D|_{\mathrm{null}(A)}$ is diagonalized by the orthonormal basis $\frac{1}{\sqrt{2n}}\bm{1}_n^T \otimes (1,0,-1,0)^T$, $\frac{1}{\sqrt{2n}}\bm{1}_n^T \otimes (0, 1,0,-1)^T$. Computing the exact proportion of diagonal perturbations $A+\epsilon D$, $\|D\|_2 = 1$, that do not preserve the sign structure of the nullspace is significantly more involved in this case, and left to the interested reader.
\end{remark}

	
	Nevertheless, we can provide sufficient conditions on a matrix $A\in\mathcal{S}(G)$ such that any nowhere-vanishhing eigenbasis of $A$ would satisfy the bounds of Theorem \ref{thm: main bounds}. For instance, Berkolaiko's rank-one argument \cite{berkolaiko2008lower}, though stated for simple eigenvalues, still holds in this case and provides an interesting consequence of multiplicity.
	
	\begin{lemma}\label{lm:gregory}
		Let $A \in \mathcal{S}(G)$ be strictly supported on $G$ with eigenvalues $$\lambda_1 \le \ldots \le \lambda_{k-1} <\lambda_k = \ldots = \lambda_{k+m-1} < \lambda_{k+m} \le \ldots \le \lambda_n.$$ Any nowhere-vanishing eigenvector $\bm{\varphi}$ in the eigenspace of $\lambda_k = \ldots = \lambda_{k+m-1}$ has
		$$ \nu(A,\bm{\varphi}) \ge k + (m-1) -1.$$
	\end{lemma}
	
	\begin{proof} Consider the spectral counting function $\lambda\mapsto N(A,\lambda)$ which equals to the number of eigenvalues of $A$ (counting with multiplicity) that are no larger than $\lambda$. We aim to show that $$ \nu(A,\bm{\varphi}) \ge  N(A,\lambda) -1,$$
 for any nowhere-vanishing $\bm\varphi\in\ker(A-\lambda I)$. 
		We proceed by induction on the first Betti number $\beta(G)$ using Berkolaiko's rank-one argument (see \cite{berkolaiko2008lower}). When $\beta(G) = 0$, $A$ is an acyclic irreducible matrix, and Fielder's matrix tree theorem states that if $\lambda$ has a nowhere-vanishing eigenvector $\bm\varphi$, then $\lambda$ is simple and $\nu(\bm{\varphi}_k) = N(A,\lambda)-1$ \cite[Corollary 2.5]{fiedler1975eigenvectors}. Now, let $\beta(G)>0 $ and suppose that our desired result holds for all connected graphs $H$ with $\beta(H)<\beta(G)$. Let $A\in\mathcal{S}(G)$ be strictly supported on $G$, $\lambda$ be an eigenvalue of $A$, possibly multiple, with a nowhere-vanishing eigenvector $\bm\varphi$. Because $\beta(G)>0$, there exists an edge $(ij) \in E(G)$ such that its removal does not disconnect $G$. Let $H$ be the graph obtained from $G$ by removing $(ij)$, so that $\beta(H)=\beta(G)-1$. We construct a rank one matrix $$R=A_{ij} \bm{\varphi}_k(i) \bm{\varphi}_k(j) \big[\mathbf{e}_i/\bm{\varphi}_k(i) - \mathbf{e}_j/\bm{\varphi}_k(j)\big] \big[\mathbf{e}_i/\bm{\varphi}_k(i) - \mathbf{e}_j/\bm{\varphi}_k(j)\big]^T,$$ such that $R\bm\varphi=0$, and the matrix $A'=A+R$ has $A'_{ij}=0$ and $A'_{rs}=A_{rs}$ for any other $r \ne s$. So $A'$ is strictly supported on the graph $H$, and $A'\bm\varphi=\lambda\bm\varphi$. Hence the induction hypothesis gives
   \begin{align}\label{eq:51}
    \nu(A,\bm\varphi)&=\nu(A',\bm\varphi)+1 \hspace{-4.5mm}&\ge& \; N(A',\lambda) \qquad &\text{if}&\quad  A_{ij} \bm{\varphi}_k(i) \bm{\varphi}_k(j)>0,\ \text{and}\\
    \nu(A,\bm\varphi)&= \nu(A',\bm\varphi) & \ge& \; N(A',\lambda)-1 \qquad &\text{if}&\quad  A_{ij} \bm{\varphi}_k(i) \bm{\varphi}_k(j)<0.\label{eq:52}
  \end{align}
 If $A_{ij} \bm{\varphi}_k(i) \bm{\varphi}_k(j)<0$ then $R$ is negative semi-definite and, by Weyl's inequality, $N(A',\lambda)=N(A+R,\lambda)\ge N(A,\lambda)$, and so \eqref{eq:52} gives $\nu(A,\bm\varphi)\ge N(A,\lambda)-1$ as needed. If, on the other hand, $A_{ij} \bm{\varphi}_k(i) \bm{\varphi}_k(j)>0$ then $R$ is rank-one and, by Weyl's inequality, $N(A',\lambda)=N(A+R,\lambda)\ge N(A,\lambda)-1$, and so \eqref{eq:51} gives $\nu(A,\bm\varphi)\ge N(A,\lambda)-1$ as needed.  
	\end{proof}

 This implies the following bounds for matrices with eigenvalue multiplicity.

 \begin{proposition}\label{prop: multiplicity}
Let $A\in \S(G)$ be strictly supported on $G$ with $r$ distinct eigenvalues of multiplicities $m_1,\ldots,m_r$, $\sum_{i=1}^r m_i = n$. Then the total nodal count, for any non-vanishing eigenbasis $\bm{\varphi}_1,\ldots, \bm{\varphi}_n$ of $A$, is bounded by
$$ {n \choose 2} + \sum_{i =1}^r {m_r \choose 2} \le  \sum_{k = 1}^n \nu(A,\bm{\varphi}_k).$$
In particular, if $\sum_{i =1}^r {m_r \choose 2} \ge \beta(G)$, then the bounds of Theorem \ref{thm: main bounds} hold for any non-vanishing eigenbasis. 
 \end{proposition}

 \begin{proof}
Let $\bm{\varphi}_1,\ldots, \bm{\varphi}_n$ be a nowhere vanishing eigenbasis of $A$, and suppose that $\lambda_{k-1}<\lambda_{k}=\ldots=\lambda_{k+m-1}<\lambda_{k+m}$. Then, by Lemma \ref{lm:gregory}, 
  $$\sum_{j=k}^{k+m-1}\left[\nu(A,\bm{\varphi}_j)-(j-1)\right]\ge  \sum_{j=k}^{k+m-1}\left[k+m-2-(j-1)\right]
  =  \sum_{i=1}^{m}(m-i)={m \choose 2},$$
  and therefore $\sum_{j = 1}^n \left[\nu(A,\bm{\varphi}_j)-(j-1)\right]\ge \sum_{i =1}^r {m_r \choose 2} $ as needed.
\end{proof}
 
We conclude that when $\sum_{i =1}^r {m_i \choose 2} \ge \beta(G)$, Theorem \ref{thm: main bounds} holds for all non-vanishing eigenbases. Next, we show that when $\sum_{i=1}^{r}{ m_{i} \choose 2}\le\beta(G)+n-1=|E(G)|$, the bounds of Theorem \ref{thm: main bounds} might still hold for all non-vanishing eigenbasis due to transversality.
	 \begin{proposition}\label{prop: trans-mult}
If $A\in \S(G)$ is strictly supported on $G$ and satisfies the transversality condition of Lemma \ref{lem: transversality}, then, for any non-vanishing eigenbasis of $A$, say $\bm{\varphi}_1,\ldots, \bm{\varphi}_n$, the total nodal count $\sum_{k = 1}^n \nu(A,\bm{\varphi}_k)$ satisfies the bounds from Theorem \ref{thm: main bounds}. Explicitly:

\begin{enumerate}
    \item The transversality condition of Lemma \ref{lem: transversality} is that 
    \[\mathcal{S}(G)+\{AX-XA:X\in\mathcal{A}(n)\}=\mathcal{S}(n).\]
    \item A necessary condition for transversality is $$\dim(\mathcal{S}(G))+\dim(\{AX-XA:X\in\mathcal{A}(n)\})
    \ge\dim(\mathcal{S}(n)).$$If $A$ has $r$ distinct eigenvalues of multiplicities $m_1,\ldots,m_r$, $\sum_{i=1}^r m_i = n$, this means that
\[|E(G)|=\beta(G)-1+n\ge \sum_{i=1}^{r}{ m_{i} \choose 2}.\] 
\end{enumerate}
 \end{proposition}
 \begin{proof}
     The statement follows directly from Lemma \ref{lem: transversality} and Theorem \ref{thm: main bounds}.
 \end{proof}
	
	\subsection{Vanishing Eigenvectors and Eigenvalue Multiplicity} We now consider the case where there is eigenvalue multiplicity and there is no non-vanishing orthogonal basis for the corresponding eigenspace. This setting is the most complex, and in general can have quite unusual behavior. The star graph is the standard example of the issues that can arise. 
	
	\begin{example}[most choices of eigenbasis do not satisfy Fiedler's matrix tree theorem]\label{ex: star}
		Let $A$ be the adjacency matrix of $G$, the star graph on $n$ vertices. The rank of $A$ is two and $A$ has nullspace
		$$\mathrm{null}(A) = \{ \bm{x} \in \mathbb{R}^{n} \, | \, \bm{x}(1) = 0, \, \bm{x}^T \bm{1} = 0\}.$$
		Any perturbation $A'$ of this matrix with simple eigenvalues and non-vanishing eigenvectors must satisfy Fiedler's matrix tree theorem \cite{fiedler1975eigenvectors}, and so must have nodal count $\nu(A',\bm{\varphi}_k) = k-1$ for all $k \in [n]$. However, we note that, of all the orthonormal bases of $\mathrm{null}(A)$, an exponentially small proportion of them have an eigenvector $\bm{\varphi}$ that, upon arbitrarily small perturbation (to, say, $\bm{\varphi}_\epsilon$), can satisfy $\nu(A',\bm{\varphi}_\epsilon)=1$ (see \cite[Lemma 4.4]{urschel2014spectral} and \cite[Figure 2]{mckenzie2023nodal} for more details). This implies that the sign structure of the large majority of eigenbases of $A$ are not achievable by any $A' \in \mathcal{S}(G)$ satisfying (NCC).
	\end{example}

\section*{Acknowledgements}
The second author thanks Mehtaab Sawhney for interesting conversions regarding diagonal perturbations. The first author was supported by the Simons Foundation Grant 601948, DJ. The authors thank Louisa Thomas for improving the style of presentation. The authors thank Adi Alon for assistance with graphic design.

{ \small 
	\bibliographystyle{plain}
	\bibliography{main.bib} }

\end{document}